\documentclass[manuscript]{acmart}
\usepackage{multirow}
\usepackage{caption}
\usepackage{subcaption}
\usepackage{graphicx}
\usepackage[utf8]{inputenc}

\AtBeginDocument{%
  \providecommand\BibTeX{{%
    \normalfont B\kern-0.5em{\scshape i\kern-0.25em b}\kern-0.8em\TeX}}}

\setcopyright{none}
\renewcommand\footnotetextcopyrightpermission[1]{}
\begin{document}

% \title{Classifying the Reader's Familiarity and Goal Conditions with Behavioral Interactions from a Mobile Device}
\title[Using Beh. Interactions to Classify the Reader's Prior Conditions]{Using Behavioral Interactions from a Mobile Device to Classify the Reader's Prior Familiarity and Goal Conditions}

%%% commented out for review %%%
\author{Sungjin Nam}
\authornote{This work was done when the author was at Adobe Research.}
\email{sjnam@umich.com}
\affiliation{%
  \institution{University of Michigan}
  \streetaddress{}
  \city{Ann Arbor}
  \state{Michigan}
  \postcode{48103}
}

\author{Zoya Bylinskii}
\email{@adobe.com}
\affiliation{%
  \institution{Adobe Research}
  \streetaddress{}
  \city{}
  \state{}
  \postcode{}
}
\author{Christopher Tensmeyer}
\email{tensmeye@adobe.com}
\affiliation{%
  \institution{Adobe Research}
  \streetaddress{}
  \city{Lehi}
  \state{Utah}
  \postcode{}
}
\author{Curtis Wigington}
\email{@adobe.com}
\affiliation{%
  \institution{Adobe Research}
  \streetaddress{}
  \city{San Jose}
  \state{California}
  \postcode{}
}
\author{Rajiv Jain}
\email{@adobe.com}
\affiliation{%
  \institution{Adobe Research}
  \streetaddress{}
  \city{}
  \state{}
  \postcode{}
}
\author{Tong Sun}
\email{tsun@adobe.com}
\affiliation{%
  \institution{Adobe Research}
  \streetaddress{}
  \city{San Jose}
  \state{California}
  \postcode{}
}

\renewcommand{\shortauthors}{Nam et al.,}

\begin{abstract}
A student reads a textbook to learn a new topic; an attorney leafs through familiar legal documents. Each reader may have a different goal for, and prior knowledge of, their reading. A mobile context, which captures interaction behavior, can provide insights about these reading conditions. 
In this paper, we focus on understanding the different reading conditions of mobile readers, as such an understanding can facilitate the design of effective personalized features for supporting mobile reading. With this motivation in mind, we analyzed the reading behaviors of 285 Mechanical Turk participants who read articles on mobile devices with different familiarity and reading goal conditions. The data was collected non-invasively, only including behavioral interactions recorded from a mobile phone in a non-laboratory setting. 
Our findings suggest that features based on touch locations can be used to distinguish among familiarity conditions, while scroll-based features and reading time features can be used to differentiate between reading goal conditions. 
Using the collected data, we built a model that can predict the reading goal condition (67.5\%) significantly more accurately than a baseline model. Our model also predicted the familiarity level (56.2\%) marginally more accurately than the baseline. 
These findings can contribute to developing an evidence-based design of reading support features for mobile reading applications. 
Furthermore, our study methodology can be easily expanded to different real-world reading environments, leaving much potential for future investigations. 
\end{abstract}
\ccsdesc[500]{Human-centered computing~Smartphones}
\ccsdesc[300]{Human-centered computing~User studies}
% \ccsdesc[100]{Human-centered computing~Ubiquitous and mobile computing design and evaluation methods}
% \ccsdesc[100]{Human-centered computing~User models}
\keywords{mobile reading, reading conditions, behavioral interactions, crowdsourcing, prediction models}

\maketitle
\pagestyle{empty}

%%%%%%%%%%%%%%%%%%%%%%%%%%%%%%%%%%%%%%%%%%%%%%%%%%%%%%%%%%%%%%%%%%%%%%%%
\section{Introduction}
Readers interact with documents in distinct ways based on various levels of expertise or prior \textit{familiarity} with a document's content.
For example, a person may read to learn about a new topic or may reread a document to refresh their memory. The reading behaviors exhibited in these two scenarios can be quite different, even for the same reader.
Readers may also differ in their \textit{reading goals}. Some may be looking for specific information, while others read to develop a deeper understanding of the topic.
These different reading goals can lead to different reading strategies and behaviors.
For information-finding, skimming the document can be more efficient. 
On the other hand, obtaining an in-depth understanding may require more careful reading and integrating information across the text.

Understanding different types of reading is a prerequisite for building personalized reading tools that can help users be more effective at their tasks % accomplish their goals 
by adapting to the content and the readers' goals. % and that adapt to their level of familiarity.
Prior studies have suggested support features for mobile reading, including varying font sizes by reading goal~\cite{wang2018evaluating}, identifying difficult sentences that have longer dwell times~\cite{oh2014generating}, and highlighting the text to restore the reader's attention and improve comprehension and engagement~\cite{mariakakis2015switchback}. 
However, these support features were limited to identifying the reader's state (e.g., struggling, distracted) when interacting with the reading material. In this study, we suggest methods that can identify the reader's initial condition going into the reading, specifically the reading goal and familiarity with the topic. 

Though reading occurs in many contexts (e.g., books, desktop displays, E-readers), our study focuses on reading in the mobile context.
Compared to other devices, such as desktop, people prefer mobile devices for particular types of reading, such as reading news articles~\cite{mitchell2017americans}. % or books~\cite{mobile2019}.
Compared to paper-based media, reading on mobile devices has advantages in portability and easy access, but disadvantages in reduced readability and ease of navigation \cite{shimray2015overview}.
Recognizing the reader's prior familiarity or reading goal conditions would be crucial for the mobile reading system to determine more effective personalized support features and improve the mobile reading experience. 

\begin{figure}[t]
  \centering
  \includegraphics[width=0.95\linewidth]{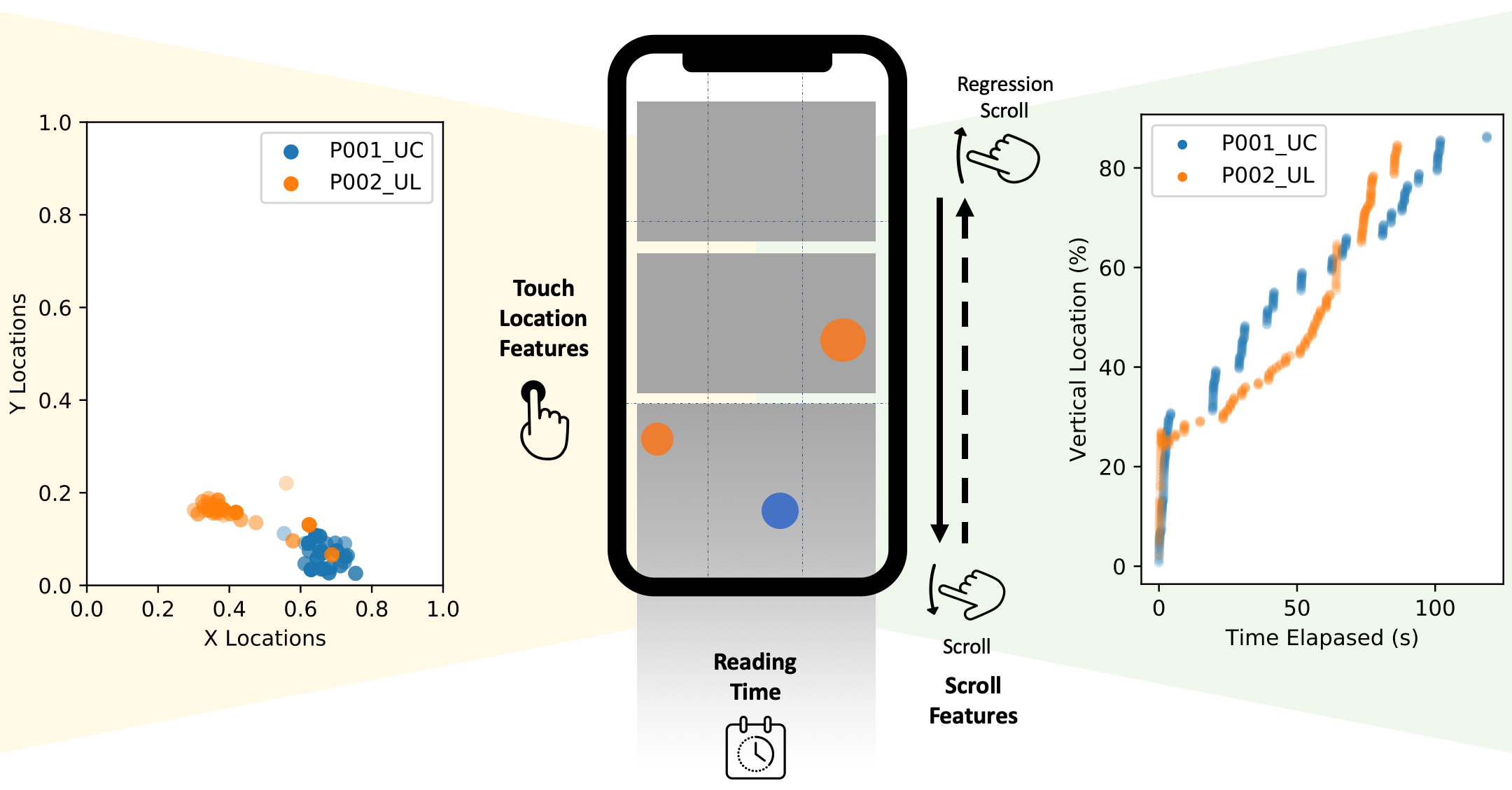}\\
  \centerline{\small{\hfill \textsf{(a) Touch Features} \hfill \textsf{(b) Mobile Device} \hfill \textsf{(c) Scroll Features} \hfill}}
  \caption{
  From the mobile device (b), we collected interaction signals related to reading time, touch features (a), and scroll features (c).
  Touch features are represented as X/Y screen coordinates. Scroll features include scroll distance, speed, direction (regular downwards scrolling, or regression scrolling backward in the document), and time spent between scrolls. 
  The colors in the graphs indicate example interactions captured from two participants (P001 and P002) that had different assigned reading goals (blue for C: \texttt{contextual} and orange for L: \texttt{literal}), but read the same (U: \texttt{unfamiliar}) article. 
  The contextual reader (P001\_UC) scrolled in bursts, while the literal reader (P002\_UL) used continuous scrolling by touching the screen more consistently in the bottom right region. 
}
  \label{fig:features_diagram}
  \Description{Interaction signal types collected from mobile devices.}
\end{figure}

In this paper, we introduce methods to capture two reading conditions: reader familiarity with the text content (\texttt{familiar} vs. \texttt{unfamiliar}) and reading goal (\texttt{literal} vs. \texttt{contextual} reading). We run our studies in a non-laboratory setting with hundreds of participants recruited through Amazon's Mechanical Turk (MTurk) to approximate reading behaviors in-the-wild on a general population pool.
Our statistical analyses of user interactions during reading on mobile devices show that the different reading conditions lead to differences in touch and scroll interaction patterns (see Fig.~\ref{fig:features_diagram}).
We also develop machine learning models that can automatically predict the reading conditions from user interactions achieving 56\% accuracy in predicting familiarity level, and 68\% accuracy in predicting reading goals. Our work opens up exciting possibilities for future reading tools to incorporate reading condition prediction to automatically adjust reading formats~\cite{baudisch2004fishnet,wallace2019fonts,chi2005scenthighlights} and personalize supporting materials~\cite{kumbruck1998hypertext,destefano2007cognitive}.

\section{Related Works}
\subsection{Distinguishing Different Types of Readers}
Readers may have different prior \textit{familiarity} with the article topic they are to read about. In previous studies, familiarity with the article's topic was considered a significant factor for the reader's engagement level~\cite{san2013towards} and comprehension level~\cite{mcnamara1996learning}. 
Other studies~\cite{destefano2007cognitive,ozuru2009prior} have suggested that more coherent text can improve reading comprehension for readers with low prior knowledge levels, while readers with more prior knowledge understand less structured and less consistent documents better. 
The reader's prior knowledge can also affect how the reader consumes and interacts with the article~\cite{o2016investigating}. 
In this paper, we chose the familiarity level as the first of our reading conditions. 
%Identifying the reader's prior knowledge level about an article is important to estimate the reader's expectation. 

One way to control a participant's familiarity is to exploit their domain expertise. 
For example, studies have used the degree program of undergraduate students to determine their familiarity with an article~\cite{mcnamara1996learning,ozuru2009prior}. 
However, this experimental setting relies on each participant's background experience, which limits the articles and domains that can be used in the experiment.
In contrast, in our study, we categorized familiarity into two conditions, \texttt{familiar} and \texttt{unfamiliar}, and manipulated the condition by controlling a participant's prior exposure to the article's contents. 
In this way, we could control the familiarity condition more easily with diverse article topics and participants. 
% more details. 

Readers may also have different \textit{reading goals} based on their circumstances. 
Studies of taxonomies in education~\cite{bloom1956taxonomy} and reading~\cite{barrett1968taxonomy} suggest there is a hierarchy of comprehension goals.
From literal comprehension to synthesizing and applying new knowledge, different reading goals can significantly affect reading behavior.
For example, if the reader's goal is to extract relevant information within a limited time frame, the reader might skim the article until she is satisfied with the findings~\cite{duggan2011skim}. 
If the reader's goal is father to build comprehensive knowledge about a topic, she may explore the article more exhaustively and try to mentally connect relevant parts together~\cite{sullivan2015learning}. 
Understanding the reader's goal would be necessary for a reading system to suggest different supporting features that can facilitate a more literal or contextual comprehension of the article. 
In this study, we simplified reading goals into two levels - \texttt{literal}: read to recognize or identify explicit information, and \texttt{contextual}: read to provoke deeper thinking, or to make a critical judgment. We controlled the reading goal condition by the instruction given to participants before they read an article.
% In this way, we built prediction models that can predict very different reading goals. 

Overall, for this paper, we use four different types of reading conditions, combining familiarity and reading goals into a 2x2 framework that can cover a wide variety of reading scenarios (Table \ref{tab:ex_scenarios}). More details about the experiment settings can be found in Section \ref{sec:reading_conditions}.

\begin{table*}
\caption{
Example reading scenarios for different familiarity and reading goal conditions. 
%Each category includes different types of reading.
% Reading condition categories and example reading scenarios in a 2x2 matrix. 
}
\label{tab:ex_scenarios}
\begin{tabular}{c p{2.45in} p{2.45in}}
\toprule
                &   Familiar    &   Unfamiliar\\
\midrule
Literal Goal    &   
    A paralegal reads a contract while searching for specific legal phrases.    &
    A student reads a textbook chapter to answer fill-in-the-blank questions on a new topic. \\
Contextual Goal &
    A researcher reads an academic paper to get new ideas related to a topic that she has expertise in. &
    %A product manager reads through user interview transcripts about a new product to identify needed features. \\
    A government official reads a report on a new proposal in order to make budgeting decisions. \\
\bottomrule
\end{tabular}
\end{table*}

\subsection{Using Implicit Signals to Understand Reading Behaviors}
Previous studies used various methods to record behavioral signals to understand the reader's cognitive state. 
For instance, attention signals, such as eye movements, have often been considered a proxy of different cognitive states \cite{mak2019mental,leon2019specific}. 
\citet{mak2019mental} showed that different types of mental simulations during reading, such as simulating motor movements vs. perceptual recognition of the paragraph just read, are related to the duration of gaze. 
\citet{leon2019specific} suggested that different types of instructions provided before reading (i.e., reading goal) can elicit different fixation and saccade patterns. 
These results show how low-level attention features are related to various cognitive states. However, these studies require the use of external sensors (e.g., eye trackers) in controlled lab settings to collect data from participants. 
In our study, we focus on behavioral features that can be collected from any mobile device and environmental setting. 

Other studies in human-computer interaction show more convenient ways to predict different user states. 
Implicit signals, such as dwell time~\cite{lagun2016understanding} or mouse movements~\cite{guo2008exploring}, were used to determine the reader's attention location or intention in desktop web browsing. 
Reading on a mobile device is very different from reading physical media or reading on a desktop display because the screen size is smaller, and the interaction methods are different~\cite{oh2014generating}. 
Prior studies have used interactions unique to mobile devices, such as scrolling behavior and touch locations, to understand reading behaviors on mobile devices for inferring different levels of comprehension and engagement~\cite{guo2018understanding}, predicting satisfaction with visual aspects of the document~\cite{wang2018evaluating}, and identifying important sentences from the article~\cite{oh2014generating}.

% People spend more time on mobile devices to read various things, including text messages, email, news articles, social media feeds, and books~\cite{shimray2015overview}. 
Mobile devices are making reading more accessible to different parts of the world, thereby increasing the total time people spend reading~\cite{west2014reading}. 
Although previous studies have investigated different cognitive states of readers on mobile devices, including engagement \cite{guo2018understanding} or satisfaction \cite{wang2018evaluating}, there were no studies that examined the relationship between implicit behavioral interactions and the reader's initial condition going into the reading. 
Understanding the reader's prior condition, such as familiarity and reading goals, can be crucial for designing a reading system with personalized features. 
In our study, we suggest experimental methods to control different reading conditions and collect behavioral interaction features from a mobile device in a non-laboratory setting. We also investigate if we can predict the reader's condition based on behavioral interaction signals that are collected non-invasively.

% As previous studies investigated different cognitive states of readers on mobile devices, including engagement \cite{guo2018understanding} or satisfaction \cite{wang2018evaluating}, it was difficult to find studies that examine the relationship between implicit behavioral signals and the reader's prior condition. 
% Our study investigates the relationship between behavioral interactions and the reading conditions. % that are related to the reader's familiarity level and reading goal. 
% We also suggest experimental methods to control different reading conditions, and captures behavioral signals from mobile devices in non-laboratory settings. 

% . For example, the size of surface is smaller and uses vertical scrolls rather than horizontal pagenation.
% Understanding reading behavior on mobile device may require different xxx, but also can lead to more flexible and personalized solution that can lead to better reading experience \cite{}.

%%%%%%%%%%%%%%%%%%%%%%%%%%%%%%%%%%%%%%%%%%%%%%%%%%%%%%%%%%%%%%%%%%%%%%%%
\section{Research Questions}
\label{sec:rqs}
Motivated by previous studies, we put forth three research questions to address in this paper. These research questions investigate the relationship between behavioral interaction signals that are recorded from mobile devices across reading conditions (i.e., \texttt{familiar} or \texttt{unfamiliar} reading, and \texttt{literal} or \texttt{contextual} goals). 
% The findings from these research questions can contribute to understanding different mobile reading conditions through behavioral interaction signals that collected non-invasively from non-laboratory settings. 

\begin{itemize}
\item RQ1: Are there differences in behavioral signals recorded from mobile devices across reading conditions?
\item RQ2: Can we use behavioral signals to predict a person's reading condition?
\item RQ3: Can we use nested experimental conditions and individual differences to boost prediction performance?
% Can accounting for individual differences improve the prediction of the reading condition?
\end{itemize}

The first research question explores how each behavioral interaction variable that we measure during reading can be distinguished by the reading conditions. Previous studies showed that behavioral interactions, such as scroll patterns or touch locations, can be used to infer various cognitive states, including satisfaction \cite{wang2018evaluating} or engagement \cite{o2016investigating} with the content. In our study, we show how each behavioral signal recorded from a mobile phone can be used to differentiate the reading conditions that are related to the reader's goal and familiarity with the topic.

The second and third research questions investigate if we can build computational models that can effectively predict different reading conditions. 
For the second research question, we develop a lasso logistic regression model with behavioral interaction predictors. The results from our feature importance analysis show which features are more important for predicting each reading condition type. 

Based on a preliminary study (Appendix \ref{app:pilots}), we observed different individual scrolling styles: frequent vs. occasional scrollers. Taking into account behavioral interactions at an individual level (i.e., incorporating individual reading styles) may therefore lead to extra insights. Motivated by this observation, the third research question focuses on developing a more sophisticated mixed-effect model that additionally includes variables for nested experimental conditions and individual differences in reading.
% Including variables for other repeating experimental conditions, such as fixed conditions or article stimuli number, may also address biases in the dataset. 

% The second research question investigates whether we can build prediction models that can effectively predict different reading conditions by using multiple behavioral interactions together. 
% By building a prediction model, we can generalize the patterns across different behavioral interactions and understand how mobile device readers behave in different reading conditions. 
% We also explore which features are most predictive of the reading conditions, to suggest data-driven evidence for designing personalized reading support features. 

% The third research question aims to include individual differences in a prediction model. 
% To address this research question, we consider a more sophisticated modeling approach. 

%%%%%%%%%%%%%%%%%%%%%%%%%%%%%%%%%%%%%%%%%%%%%%%%%%%%%%%%%%%%%%%%%%%%%%%%
\section{Reading Conditions and Behavioral Interactions}
For our study, we defined different reading conditions to represent the reader's prior \textit{familiarity} and \textit{reading goals}. 
We then extracted various behavioral features based on interaction signals to analyze and predict these reading conditions. 
This section describes how all these variables are operationalized and recorded by the application used in our experiments. 
% For the study, we developed a web-based application that the participant can access with a mobile phone. The application assigned different experiment conditions to participants, and recorded various features related to reading, such as behavioral interactions and device metadata. 
% These features were used in different parts of the paper, including filtering participants' data and analyzing the results of preliminary and main studies. 

\subsection{Reading Conditions}
\label{sec:reading_conditions}
\begin{figure}[h]
  \centering
%   \includegraphics[height=2.6in]{figures/cond_familiar.png}
%   \hfill
%   \includegraphics[height=2.6in]{figures/cond_goal.png}
  \includegraphics[width=1.0\linewidth]{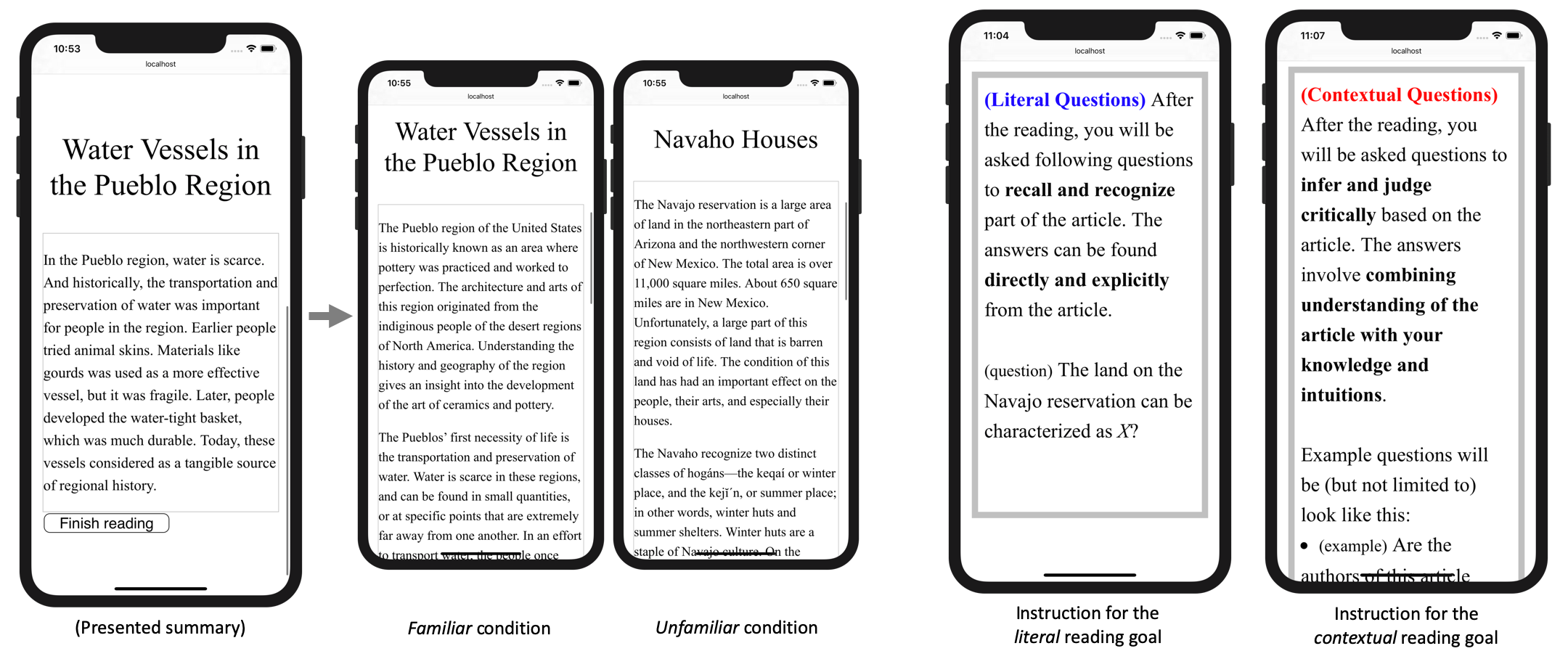}\\
  \centerline{\small{\hfill \textsf{(a) Familiarity Condition} \hfill \textsf{(b) Reading Goal Condition} \hspace{0.5in}}}
  \caption{
  The familiarity condition (a) was controlled by exposing participants, during the practice phase, to a summary of the article they would read later. In the unfamiliar condition, a summary of an unrelated article would be presented instead.
  The reading goal condition (b) was controlled by changing the wording of the instructions and sample questions provided before the article.
  }
  \label{fig:cond_manipulate}
  \Description{Assigning familiarity and reading goal conditions}
\end{figure}

In this study, we used a 2x2 framework to capture the reader's familiarity levels (\texttt{familiar} vs. \texttt{unfamiliar}) and reading goals (\texttt{literal} vs. \texttt{contextual}). Figure \ref{fig:cond_manipulate} shows how we manipulated the familiarity levels and reading goal conditions in our experiments.
All participants completed a practice session at the beginning of the experiment, whereby they read a short reading passage, corresponding to an article summary. We used this practice session to manipulate the familiarity condition for a given study. Specifically, we either had the summary correspond to an article the participant would read in the main study (\texttt{familiar} condition) or be about a different article altogether (\texttt{unfamiliar} condition).
%First, we used prior exposure to the related content to control the familiarity level. For example, we considered the \texttt{familiar} condition as the participant previously seen the related summary from the practice session before she reads the full article in the experiment session. For the \texttt{unfamiliar} condition, the participant was not exposed with the corresponding summary for the article in the experiment session.

To control the reading goal condition, we presented different instructions before the article. 
The instructions informed participants that two comprehension questions would be presented after reading. Each question was multiple-choice with three options.
For the \texttt{literal} reading condition, participants were told that the answers to the comprehension questions could be found ``directly and explicitly from the article''. Also, the exact question phrasings were provided to encourage the literal search process during reading. 
On the other hand, for the \texttt{contextual} reading condition, participants were told that answering the questions would ``involve combining an understanding of the article with your knowledge and intuitions''. Sample question phrasings were provided (as guidance), but were not identical to the questions asked at the end of the reading.

\subsection{Experiment Application and Behavioral Interactions}
We used a web-based application to show different articles and comprehension questions to participants.
While a participant read articles inside the application on a mobile device, the application also collected data, including meta-data about the accessing device and behavioral signals based on scroll and touch interactions during reading.
%Our application recorded various features related to reading, including behavioral interactions and device metadata. 
The data features collected were used in different ways, including for filtering participant data and for analyzing the results of the preliminary and main studies. 

\subsubsection{Browser Information}
\label{sec:feat_meta}
As the participant's mobile phone loaded the web-page for reading, our application recorded the browser's metadata, including the browser's name, the screen size (width and height) of the device, and each article's width and height as rendered on the device.
As described later in Section \ref{sec:mturk_participants}, this data was useful to confirm the authenticity of data collected from mobile phones. 
It also included information about how each article was rendered on different screens, which was essential to track which paragraphs were displayed on the screen and compare recorded features from different devices.

\subsubsection{Raw Interaction Signals}
\label{sec:feat_raw}
A benefit of using interaction signals as features is that they can be collected non-invasively from a mobile phone, without using external sensors like eye tracking devices. 
Our application recorded various signals during reading, including timestamps, the vertical location of the scroll bar, and X/Y touch locations. 
Data was recorded every time the vertical location of the scroll bar changed by more than 1\% of the accessing device's screen height. When a participant finished reading an article, data was saved as a JSON file in the application's server. 
% Data like timestamps, vertical location of the scroll bar, and touch locations were recorded during the read. 

\subsubsection{Extracting Predictive Features}
\label{sec:feat_pred}
Based on the raw interaction signals recorded, we extracted features that correspond to more meaningful information about reading behavior. Specifically, we extracted four types of behavioral interaction features: \textit{Reading Time}, \textit{Scroll}, \textit{Regression Scroll}, and \textit{Touch Location}. All extracted features were log-transformed and standardized for the analysis. The full list of features are listed in Table \ref{tab:anova2}. 

\textit{Reading Time} features record how much time a participant spent reading an article. For the analysis, we calculated the total time spent on the article and (average and standard deviation of) time spent per paragraph, except for the instructions. All reading time features were normalized by the heights of article paragraphs when rendered on each participant's device.

\textit{Scroll} features include frequency, travel distance, and speed of each scroll sequence. 
We considered a set of vertical locations of the scroll bar as a single scroll sequence if they share the same touch location that initiated the scrolling. Scroll information can be interpreted as a proxy for the participant's attention, as it provides information about the participant's view port per unit time, as the reading progresses \cite{oh2014generating}. 

\textit{Regression Scroll} features are similar to \textit{Scroll}, but only consider scroll sequences in the opposite direction to the text flow. This corresponds to points during reading where a participant revisits earlier parts of the article. 

\textit{Touch Location} features measure the relative distribution of X/Y locations for touch inputs. 
We arbitrarily divided the device's screen into three horizontal (left, mid, and right) and three vertical (low, mid, and high) locations, and counted the frequencies of touch interactions in each location block.
To disregard the differences between participants' dominant hands, we grouped off-center (left and right) inputs together and compared them in frequency to central (mid-horizontal) touch locations. Frequency information per location was recorded as a fraction of the total number of touch interactions. 
% Touch locations can provide additional information on the participant's search process during reading. 
% The higher the touch location is, the amount of readable text area is smaller during the scrolling. Consistency of touch location may also indicate meaningful information about the reader's comprehension state \cite{wang2018evaluating}.

\begin{table*}
\caption{
The list of extracted behavioral interaction features. 
Based on a two-way ANOVA analysis, \textit{Reading Time} and \textit{Scroll} features were significantly different across reading goal conditions (\texttt{contextual} vs. \texttt{literal} reading).
\textit{Touch} features were significantly different across familiarity level conditions (\texttt{familiar} vs. \texttt{unfamiliar}). 
($^{.}$: $p < 0.1$, $^{*}$: $p < 0.05$, $^{**}$: $p < 0.01$, $^{***}$: $p < 0.001$)}
\label{tab:anova2}
\begin{tabular}{r l l l}
\toprule
% Feature Type        &   Feature Name                & \multicolumn{2}{c}{two-way ANOVA}\\
Feature Type        &   Feature Name                                                       & F vs. U       & C vs. L \\
\midrule
Reading Time        &   Total article read time                                            &               & C < L $^{**}$                         \\ % &   &$^{*}$   \\
                    &   Read time per paragraph (Avg., Std.)                               & F < U $^{.}$ (Std.) & interaction $^{.}$ (Avg.)               \\ % &   &   \\
Scroll              &   Scroll frequency                                                   &               &                                       \\ % &   &   \\
                    &   Time between scroll sequences (Avg., Std.)                          &               &                                       \\ % &   &   \\
                    &   Travel distance per scroll sequence (Avg., Std.)                   &               & C < L $^{.}$ (Avg.), $^{***}$(Std.)   \\ % &   &   \\
                    &   Scrolling speed per sequence (Avg., Std.)                          &               & C < L $^{**}$ (Avg.), $^{***}$(Std.)  \\ % &   &   \\
Regression Scroll   &   Scroll frequency                                                   &               & C < L $^{**}$, interaction $^{.}$     \\ % &   &   \\
                    &   Travel distance per scroll sequence (Avg., Std.)                   &               & C < L $^{**}$ (Avg.)                  \\ % &   &   \\
                    &   Scrolling speed per sequence (Avg., Std.)                          &               & C < L $^{*}$ (Avg.), $^{***}$(Std.)   \\ % &   &   \\
                    &   Revisiting the instruction paragraph                               &               & C < L $^{***}$                        \\ % & $^{.}$  &$^{.}$   \\
Touch Location      &   X-Axis: Ratio of touch initialization from the left or right sides &               &                                       \\ % &   &   \\
                    &   X-Axis: Std. of left/mid/right touch ratios                        &               &                                       \\ % &   &   \\
                    &   Y-Axis: Ratio of touch initialization from the low screen          & F > U $^{**}$ &                                       \\ % &$^{.}$   &   \\
                    &   Y-Axis: Ratio of touch initialization from the mid screen          & F < U $^{**}$ &                                       \\ % &$^{.}$   &   \\
                    &   Y-Axis: Ratio of touch initialization from the high screen         &               &                                       \\ % &   &   \\
                    &   Y-Axis: Std. of left/mid/high touch ratios                         & F > U $^{***}$&                                       \\ % &   &   \\
% TODO: add visual icons per feature type
\bottomrule
\end{tabular}
\end{table*}

%%%%%%%%%%%%%%%%%%%%%%%%%%%%%%%%%%%%%%%%%%%%%%%%%%%%%%%%%%%%%%%%%%%%%%%%
\section{Main Study Design}
Based on our initial investigations, we sought to launch large-scale crowdsourcing studies to investigate how participant reading behaviors, measured by reading time, scroll, and touch features, differ by familiarity and reading goal conditions. 
% reading goals and familiarity with the topic read. 

\subsection{Pre-screening Questionnaire}
\label{sec:pre_screener_q}
For the main study, we recruited participants from Amazon's Mechanical Turk (MTurk). 
First, we developed a pre-screening questionnaire that would serve as a qualification task for the main study. The questionnaire collected information about a participant's native language, demographic information (e.g., age group, education level), and daily reading related behaviors (e.g., frequency of reading for work and leisure, types of reading done, reading device usage). Participants' geographic locations were limited to native English speaking countries.
We collected 1,129 pre-screening responses. All participants were paid \$0.30 for their responses. More details of the questionnaire responses can be found in Appendix \ref{app:prescreener}.

\subsection{Experiment flow}
We conducted two experiments (\textit{Experiment 1} and \textit{Experiment 2}). Both experiments shared the same experimental settings, while utilizing different articles and comprehension questions. In total, we collected data on four different articles. 
An English reading specialist curated and shortened articles from project Gutenberg\footnote{\url{https://www.gutenberg.org}} and composed comprehension questions, norming them at an eighth-grade level. We used four history articles for our experiments: Experiment 1: \textit{Manhattan in the Year 1609} (533 words), \textit{The Beginning of American Railroads} (456 words); Experiment 2: \textit{Water Vessels in the Pueblo Region} (453 words), \textit{Navaho Houses} (465 words).
% \footnote{
% The first two articles were used for the experiment session of Experiment 1, and later two articles were used for Experiment 2.
% \begin{itemize}
%     \item Manhattan in the Year 1609 \url{https://www.gutenberg.org/files/13842/13842-h/13842-h.htm\#CHAPTER_I},
%     \item The Beginning of American Railroads\url{https://www.gutenberg.org/files/3036/3036-h/3036-h.htm}
%     \item Water Vessels in the Pueblo Region \url{https://www.gutenberg.org/files/17170/17170-h/17170-h.htm\#fig490}
%     \item Navaho Houses \url{https://www.gutenberg.org/files/18206/18206-h/18206-h.htm}
% \end{itemize}
% }.

An experiment consisted of three parts: an initial survey, practice session, and test session. 
The survey asked questions about the device used for participation, the time of participation, and to briefly describe the current environment (time of day, type of light, and name of the mobile device). 
Participants were instructed not to use other plug-ins for reading
or the browser's back button during the task to prevent navigation between the article and comprehension questions (the application also was explicitly programmed to prevent this with a warning). Participants were told to take as much time as needed for reading. 
% Also, they were informed that the study does not measure their intelligence or fluency with technology devices. They could quit anytime during participation if they felt uncomfortable.

% summaries: 88, 65, 62, 48 words (65 avg.)
After the survey, participants were guided to the practice session.
In the practice session, participants became familiar with the task by reading a summary paragraph (about 65 words long) and answering comprehension questions. The practice session provided two sets of summary paragraphs and comprehension questions. 
In the test session, participants were also presented with two sets of articles and comprehension questions. The test articles were about 500 words long. 
At the end of the study, participants received a code to enter on the original study page to receive compensation. The study took approximately 11 minutes.
Participants received \$3.00 USD, and a bonus of \$0.50 if they successfully followed the instructions. 
% In the practice session, participants could get familiar with the task by reading a single paragraph (about 65 words long) and answering sample comprehension questions. The practice session provided two sets of summary paragraphs and comprehension questions.
% read an article (about 500 words long) and answered comprehension questions. The test session was also consisted of two sets. 
% In the test session, participants read two articles (about 500 words each) and answered comprehension questions after each article. 
%As the participant reached the end of the task, the application showed a thank you message with a randomly generated numeric code. Participants were instructed to copy this code and paste it back to the instruction page to finalize the task and claim the compensation.

% "Reading passages", "Participants", "UI", etc.
\subsection{Participants and Data Filtering}
\label{sec:mturk_participants}
Based on the pre-screening questionnaire, we selected MTurk workers whose native language is English and who reside in English-speaking countries. We recruited workers with HIT rates above 98\%, and more than 1000 successful HIT records. Participants who fit these criteria and accepted our HIT were provided with a URL to open on a mobile device.

Since our study involved a relatively complex reading task on a mobile device, we carefully selected data for analysis (Appendix \ref{app:filtering}). 
Initially, 372 MTurk workers (\textit{Experiment 1}: 181, \textit{Experiment 2}: 191) began our task, and 342 of them (\textit{Experiment 1}: 170, \textit{Experiment 2}: 172) completed it.
Among those who completed the task, we filtered out participants who did not properly follow the instructions, as detailed below.

The first filtering criterion was reading time (\textit{R. Time}). We filtered out participants who spent too little (less than 30 seconds) or too much (more than 300 seconds) time on any article.
Traditional reading speed studies suggest that the average reading speed for native English speakers is around 250 words per minute \cite{fry1963teaching,nation2009reading}. As our experimental conditions may be a little different than the previous studies, we set more generous ranges to include reading speeds that are two times faster or slower than regular reading. 

The second criterion was the accessing device's name (\textit{Dev. Name}). We wanted to keep this study strictly on mobile. Based on the browser metadata that we collected (Section \ref{sec:feat_meta}), we filtered out participants' data if the browser agent information did not contain the string ``iPhone'' or ``Android''. %By this way, we could rule out data generated from other devices like PCs or tablets. 

The third criterion was the window size ratio (\textit{Wndw. Size}). We filtered out data if the article was read in landscape orientation, or with other irregular screen size settings. 

The fourth criterion was if behavioral interactions were successfully recorded (\textit{Input Rec.}).
Missing records happened if the participant's device was too old to be compatible with our experiment application, or the participant read the article without touch interactions (e.g., using a desktop browser with a mobile browser profile).
%There were some participants' data that contained metadata of mobile browsers, but without any recorded interactions. From our observation, 
% was using a desktop browser's developer tool to simulate the mobile browser. 

As a result, we ended up using 285 participants' data (\textit{Experiment 1}: 141 participants, 77.90\% of recruited participants; \textit{Experiment 2}: 144 participants, 75.39\%) for the analysis. Note that we did not use the comprehension questions as a filtering criterion (more details can be found in Appendix~\ref{app:qna_response}).

%%%%%%%%%%%%%%%%%%%%%%%%%%%%%%%%%%%%%%%%%%%%%%%%%%%%%%%%%%%%%%%%%%%%%%%%

\section{Main Study Results}
\subsection{Post-Experiment Survey}
\label{sec:post_ex_survey}
For a subset of participants (52 participants from \textit{Experiment 1}, 60 from \textit{Experiment 2}), we asked some follow-up questions right after the experiment. The questionnaire asked participants to reflect on the different reading conditions that they were assigned to, as well as whether and how they read differently per condition.
% The actual questionnaire can be found in Appendix xxx. 

The first two questions asked about the summaries provided in the practice session. Of the participants who read at least one \texttt{familiar} condition article (read a summary in the practice session that corresponded to an article in the test session), %72\%), 
many reported (74\%) that the summary gave them a quick idea of the longer article, made the main article easier to skim, and helped them remember the details better. These results indicate that participants made use of the fact that the summary familiarized them with the topic of the article.%were aware that the content of summary paragraphs corresponded to the full articles in the familiarity condition. 

The next two survey questions asked about the reading goals that were assigned via the instruction wording. 
For these questions, more participants reported (58\%) that the instruction did not affect their reading (i.e., that they read normally), and did not use any special strategies given the relatively short article lengths. 
These results might imply that our instructions may not have been effective in guiding participants towards a particular goal, or it could indicate that introspective self-assessment of reading behavior is difficult.
%These were mixed results on our use of instruction to controll the reading goal conditions may not worked for all participants. 
The last survey question asked about the perceived difficulty of the comprehension questions. Most participants reported that the difficulty was OK (89\%), and only a few participants found the questions too easy (11\%). No participants reported that the questions were too hard. %Participants could also provide any additional comments about the study.

\subsection{Comparing Observed Features between Reading Conditions}
\label{sec:res_anova}
Recall RQ1: \textit{Are there differences in behavioral signals recorded from mobile devices across reading conditions?}

To answer this research question, we conducted a two-way ANOVA analysis comparing which behavioral features (including interactions) differ by either familiarity level or reading goal (Table \ref{tab:anova2}).
% From the TukeyHSD post-hoc analysis, we could find that observations from Reading Time, Scroll, and Regression Scroll features show differences between reading goal conditions. 
For the familiarity level (F vs. U), we found that participants reading \texttt{familiar} articles tended to touch the lower part of the screen more ($p < 0.01$), and were less consistent (i.e., higher standard deviation) in their vertical touch locations ($p < 0.001$). 
% This contradicted results from our pilot study (Section~\ref{sec:preliminary_res}) and a prior study~\cite{wang2018evaluating}, both of which suggested that the touch locations of readers in the \texttt{familiar} condition tend to be more consistent. 
We did not find any significant interactions between the familiarity and reading goal conditions. 

Across reading goal conditions (C vs. L), \texttt{contextual} reading corresponded to more linear reading behaviors. 
With the \texttt{contextual} reading goal, scrolling distance was more consistent (i.e., lower standard deviation; $p < 0.001$), and scrolling speed was slower ($p < 0.01$) and more consistent ($p < 0.001$) than for the \texttt{literal} reading goal. 
The \texttt{contextual} reading goal also correlated to less reading time ($p < 0.01$), fewer regression scrolls (for revisiting previous paragraphs ($p < 0.01$) and instructions ($p < 0.001$)), longer regression scrolling distance ($p < 0.001$), and slower ($p < 0.05$) and more consistent regression scrolling speed ($p < 0.001$). 
These results can be interpreted to mean that the \texttt{contextual} reading goal elicited more linear navigation during reading. Regression scrolls were less frequent (but longer if needed), eventually leading to a shorter overall reading time. 
% These results match our pilot study results (Section \ref{sec:preliminary_res}).

We also conducted a 1-way mixed ANOVA analysis for each familiarity condition and reading goal separately. This analysis includes repeated measures in the analysis, such as the fixed-effect conditions of each participant, to more correctly reflect our study design. However, we found no features to be significantly different across conditions in this analysis (Appendix \ref{app:anova_mixed}).

In summary, we found that touch location-based features can effectively distinguish between the familiarity conditions. Features based on scrolling and reading time were also effective for differentiating the reading goal conditions.

\subsection{Predicting Reading Conditions}
In the previous section we identified differences in behavioral interactions across reading conditions.
To answer the second and the third research questions, we developed prediction models to investigate if these behavioral interactions can be used together to predict each article's familiarity level or reading goal conditions. 

\vspace{\baselineskip}

Recall RQ2: \textit{Can we use behavioral signals to predict a person's reading condition?}

\subsubsection{Non-Mixed-Effects Modeling}

We trained Logistic Regression models with an L1 penalizing term (Lasso-LR) to predict either the familiarity level or the reading goal individually.
The penalizing term tries to prevent the model from over-fitting by encouraging multiple feature coefficients to be zero. 
We selected the L1 penalizing term of each Lasso-LR model through the nested cross-validation process. First, we divided the data into ten outer-folds.
Second, from each outer fold, we randomly divided the training set into ten inner-folds, and selected the L1 penalizing term that provided better prediction accuracy scores across ten validation sets. 
Finally, we selected the L1 term that performed the best across the ten outer-folds. 
% We first used all of the data to estimate a reasonable value for the L1 penalizing term using 10-fold cross validation.
% N-fold cross validation is performed by randomly dividing the data into N parts and training N models on different subsets of N-1 parts, where each model is evaluated on its held-out \nth{N} part.
After we decided on the L1 penalizing term, we conducted the regular 10-fold cross-validation for the same dataset to get the prediction results on each fold's test set. 

\begin{table*}
\caption{
Average performance of predicting the familiarity level (F vs. U) and reading goal (C vs. L) conditions with Lasso-LR and GLMER models. 
Recall, precision, and F1 scores are reported as the average of negative (N) and positive (P) labels. 
Also, these scores for the baseline models are omitted since this model always predicted the same label. 
The numbers in parentheses are 95\% confidence intervals.
The best prediction score, if significantly better than the baseline (or another model), is marked in bold. Scores only marginally better than the baseline (or another model) are italicized.
% The best prediction score, if significantly better than the other model, is marked in bold. Scores only marginally better than the other model are italicized.
}
\label{tab:models_preds}

% \begin{tabular}{r l l l  l l l  l l l}
% \toprule
%                 & \multicolumn{3}{c}{F vs. U}                                                           & \multicolumn{3}{c}{C vs. L}\\
%                 & Baseline              & Lasso-LR                      & GLMER                         & Baseline              & Lasso-LR                      & GLMER                         \\ 
% \midrule
% Accuracy        & 0.511 (0.468, 0.555)  & 0.523 (0.484, 0.562)          & \textit{0.556 (0.526, 0.586)} & 0.517 (0.471, 0.564)  & \textbf{0.702 (0.678, 0.726)}  &  0.670 (0.627, 0.714)\\
% Recall (N+P)    & -                     & \textit{0.529 (0.492, 0.565)} & 0.472 (0.443, 0.501)          & -                     & \textit{0.700 (0.675, 0.725)}  &  0.662 (0.620, 0.704)\\
% Precision (N+P) & -                     & \textit{0.529 (0.485, 0.572)} & 0.453 (0.399, 0.507)          & -                     & \textit{0.711 (0.686, 0.735)}  &  0.707 (0.658, 0.756)\\
% F1 (N+P)        & -                     & \textbf{0.513 (0.471, 0.555)} & 0.391 (0.363, 0.419)          & -                     & \textit{0.696 (0.670, 0.722)}  &  0.647 (0.599, 0.696)\\
% % TODO: shrink/split; color columns by different reporting types
% \bottomrule
% \end{tabular}

\begin{tabular}{r r l l l}
\toprule
                        &                       & Baseline              & Lasso-LR                       & GLMER                         \\ 
\midrule
\multirow{4}{*}{F vs. U}    & Accuracy          & 0.511 (0.472, 0.549)  & \textit{0.562 (0.505, 0.618)}  & 0.556 (0.530, 0.582)\\
                            & Recall (N+P)      & -                     & \textit{0.560 (0.505, 0.616)}  & 0.476 (0.431, 0.520)\\
                            & Precision (N+P)   & -                     & \textit{0.561 (0.503, 0.618)}  & 0.462 (0.358, 0.567)\\
                            & F1 (N+P)          & -                     & \textbf{0.553 (0.494, 0.611)}  & 0.390 (0.348, 0.432)\\\hline
\multirow{4}{*}{C vs. L}    & Accuracy          & 0.519 (0.501, 0.538)  & \textbf{0.675 (0.629, 0.721)}  & 0.672 (0.646, 0.699)\\
                            & Recall (N+P)      & -                     & \textit{0.673 (0.625, 0.721)}  & 0.663 (0.633, 0.694)\\
                            & Precision (N+P)   & -                     & 0.680 (0.631, 0.728)           & \textit{0.709 (0.661, 0.757)}\\
                            & F1 (N+P)          & -                     & \textit{0.670 (0.622, 0.718)}  & 0.651 (0.620, 0.681)\\
\bottomrule
\end{tabular}
\end{table*}

Table \ref{tab:models_preds} shows the average prediction performance of the baseline and the Lasso-LR models across 10-folds. %(on the held-out test data, averaged across 10 train/test folds). 
The baseline models represent the prediction performance without using any behavioral interaction features (i.e., predicting the majority label).
For predicting the familiarity level (F vs. U), our Lasso-LR model performed marginally better than the baseline accuracy. 
For predicting the reading goal (C vs. L), the Lasso-LR model performed significantly better than the baseline accuracy. 

\newcommand\tbclw{1.25in} %{1.12in}
\begin{table*}
\caption{
Lasso-LR and GLMER model ablation experiments: average differences in prediction performance when each feature is removed from the corresponding Lasso-LR model. 
The smaller the value, the more the feature contributed to the full model.
Recall, precision, and F1 scores are reported as the average of negative (N) and positive (P) labels. 
The numbers in parentheses are 95\% confidence intervals. 
The contributions that are significantly different from 0 are marked in bold. If there are no significant contributions found, the best marginal contributions per score type are italicized.
% Some scores for the Lasso-LR model are omitted if the model did not include the feature type by the selected penalized term. 
Some scores for the Lasso-LR model are omitted if coefficients are estimated as zero with the selected penalized term. 
}
\label{tab:feat_importance}
\begin{tabular}{p{0.3in} p{0.63in}  p{\tbclw} p{\tbclw} p{\tbclw} p{\tbclw}}
\toprule
\multicolumn{2}{l}{Lasso-LR}                & $-$ Reading Time          & $-$ Scroll                & $-$ Regression Scroll     & $-$ Touch Location   \\
\midrule
\multirow{4}{*}{F vs. U}&Accuracy           &   -  &   -  &   -  &   \textbf{-0.068 (-0.116, -0.020)} \\
                        &Recall (N+P)       &   -  &   -  &   -  &   \textbf{-0.073 (-0.117, -0.030)} \\
                        &Prec. (N+P)        &   -  &   -  &   -  &   \textbf{-0.246 (-0.353, -0.140)} \\
                        &F1 (N+P)           &   -  &   -  &   -  &   \textbf{-0.182 (-0.262, -0.101)} \\ \hline
\multirow{4}{*}{C vs. L}&Accuracy           &   \textbf{-0.109 (-0.154, -0.064)} &   -0.029 (-0.076, 0.017)  &   -0.000 (-0.035, 0.035)    &   0.002 (-0.036, 0.040) \\
                        &Recall (N+P)       &   \textbf{-0.111 (-0.157, -0.066)} &   -0.028 (-0.077, 0.020)  &   -0.000 (-0.036, 0.035)    &   0.001 (-0.036, 0.038) \\
                        &Prec. (N+P)        &   \textbf{-0.111 (-0.160, -0.062)} &   -0.024 (-0.071, 0.023)  &   -0.000 (-0.035, 0.035)    &   0.003 (-0.037, 0.044) \\
                        &F1 (N+P)           &   \textbf{-0.119 (-0.165, -0.074)} &   -0.031 (-0.082, 0.020)  &   -0.000 (-0.036, 0.035)    &   0.001 (-0.036, 0.038) \\
% TODO: shrink/split; color columns by different reporting types
\bottomrule
\end{tabular}

\begin{tabular}{p{0.3in} p{0.7in}  p{\tbclw} p{\tbclw} p{\tbclw} p{\tbclw}}

\toprule
\multicolumn{2}{l}{GLMER}            & $-$ Reading Time          & $-$ Scroll                & $-$ Regression Scroll     & $-$ Touch Location   \\
\midrule
\multirow{4}{*}{F vs. U}&Accuracy           &   -0.007 (-0.031, 0.016)  & \textbf{-0.018 (-0.031, -0.004)}  &    0.011 (-0.014, 0.035)           &   -0.011 (-0.035, 0.013)\\
                        &Recall (N+P)       &    0.004 (-0.022, 0.030)  & -0.002 (-0.028,  0.024)           &   \textit{-0.015 (-0.057, 0.026)}  &    0.020 (-0.020, 0.059)\\
                        &Prec. (N+P)        &   -0.008 (-0.063, 0.047)  & -0.012 (-0.080,  0.056)           &   \textit{-0.064 (-0.142, 0.014)}  &    0.007 (-0.091, 0.105)\\
                        &F1 (N+P)           &   -0.009 (-0.038, 0.020)  & -0.012 (-0.044,  0.019)           &   \textit{-0.028 (-0.070, 0.014)}  &   -0.006 (-0.044, 0.032)\\ \hline
\multirow{4}{*}{C vs. L}&Accuracy           &   \textbf{-0.097 (-0.136, -0.058)} & \textbf{-0.051 (-0.086, -0.017)} &   -0.018 (-0.047, 0.011)  & 0.002 (-0.028, 0.031)\\
                        &Recall (N+P)       &   \textbf{-0.105 (-0.146, -0.064)} & \textbf{-0.052 (-0.088, -0.017)} &   -0.018 (-0.047, 0.011)  & 0.002 (-0.028, 0.032)\\
                        &Prec. (N+P)        &   \textbf{-0.102 (-0.167, -0.037)} & -0.045 (-0.104,  0.015)          &   -0.013 (-0.051, 0.024)  & 0.001 (-0.035, 0.036)\\
                        &F1 (N+P)           &   \textbf{-0.137 (-0.179, -0.094)} & \textbf{-0.067 (-0.112, -0.022)} &   -0.023 (-0.053, 0.007)  & 0.003 (-0.029, 0.034)\\
% TODO: to be filled; shrink/split; color columns by different reporting types
\bottomrule
\end{tabular}
\end{table*}

Table \ref{tab:feat_importance} includes a more detailed analysis of each feature's contribution to the corresponding Lasso-LR model. 
For predicting the familiarity level (F vs. U), touch location features were more important than other features in contributing to the Lasso-LR model's accuracy, precision, and F1 scores. 
These results match the ANOVA results from Section \ref{sec:res_anova}, showing that touch location features are closely related to a reader's familiarity levels with the reading topic.

For predicting the reading goals (C vs. L), reading time was the most important feature type, significantly contributing to the model's accuracy, precision, and F1 scores. Scroll-based features also marginally contributed to the model's recall score. The results for predicting the reading goal conditions were slightly different from the ANOVA results from Section \ref{sec:res_anova}, which showed that many scroll-based features were significantly different between the \texttt{contextual} and \texttt{literal} reading goal conditions. 
Touch location features were consistently less important features in both ANOVA and Lasso-LR models. 

\vspace{\baselineskip}

Recall RQ3: \textit{Can we use nested experimental conditions and individual differences to boost the prediction performance?}

\subsubsection{Mixed-Effects Modeling}
\label{sec:res_glmer}
From the pilot study, we observed that individual readers might have distinctive scrolling styles when reading on a mobile device. For example, some engage in frequent scrolling, while others only occasionally scroll during reading (Figure \ref{fig:features_diagram}, Appendix \ref{app:pilots}). 
Motivated by these observations, we developed a mixed-effects logistic regression model (GLMER) to predict reading conditions and address the third research question. The GLMER models are expected to capture biases that may be caused by these repeated measures from the experimental settings, and to account for individual scrolling styles when using scroll-based features for prediction. 
We used the \textit{lme4} package \cite{lme4} to build the GLMER models\footnote{We set the parameter \texttt{nAGQ=0} for GLMER models to help with convergence.}.

\begin{figure}[h]
  \centering
  \includegraphics[width=\linewidth]{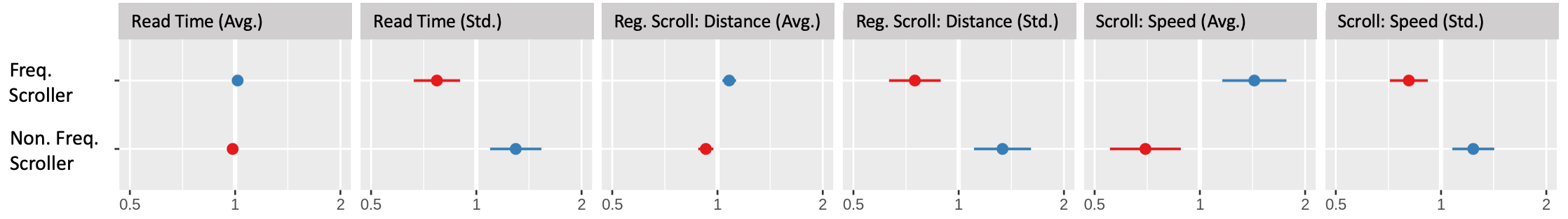}
  \caption{
  % Random slope results from the GLMER model for predicting the reading goal conditions (C vs. L). 
  Each plot shows how the random slope coefficients for the behavioral features from the GLMER model are negatively (red) or positively (blue) related to the \texttt{literal} reading condition, given by the participant's reading style (frequent or infrequent scroller).
  The numbers on the x-axis are log odds. 
  }
  \label{fig:ranef_slope}
  \Description{Interaction signal types collected from mobile devices.}
%   TODO: update feature recording plots
\end{figure}

Figure \ref{fig:ranef_slope} shows how the GLMER model interpreted behavioral signals of readers with different scrolling styles, even if they have the same reading goal condition. We selected the features from each feature type with above marginal contributions for predicting the reading goal condition (Table \ref{tab:feat_importance}).
For example, the plot illustrates that if you are a frequent scroller (i.e., generate more scroll sequences than the median), the \texttt{literal} reading goal may evoke more consistent reading time per paragraph, a longer and more inconsistent regression scroll distance, and faster but inconsistent scrolling speed. 
However, if you are an infrequent scroller, the same \texttt{literal} reading goal may lead to opposite behavioral patterns. 

%Also, as mentioned in Section \ref{sec:reading_conditions}, each participant read two articles while one of the reading condition factors (either the familiarity level or the reading goal) was fixed. 

% evoke a faster scrolling speed, with longer, regular stops between scrolls. 
% However, if you are not a frequent scroller, the same \texttt{literal} reading goal may cause a higher inconsistency in scrolling speed and time spent between between scrolls. 

For the GLMER model, scroll and regression scroll features are the more important features for predicting the familiarity levels. 
For predicting the reading goal, reading time and scroll features were the most important features (Table \ref{tab:feat_importance}). 
However, Table \ref{tab:models_preds} shows that the GLMER model did not improve upon the Lasso-LR model in predicting the familiarity and reading goal conditions. Although the model can capture various behavioral patterns by individual scrolling styles, further investigation on how behavioral features interact within the different reading conditions, and more thorough feature selection process for better model convergence may be required to start to see prediction performance benefits.

% we used following techniques to improve the model's stability. We used \texttt{nAGQ=0} parameter for GLMER models. Using this parameter may derive less accurate coefficient estimations, but helped model convergence. Our GLMER models used fewer predictive features than the Lasso-LR models. 
% Using the variance inflation factor (VIF) score, we iteratively removed predictive features until the model converged without error, and avoided the multi-colinearity problem.
%}. 
% but slightly worse than the Lasso-LR model results in predicting the familiarity and the reading goal conditions. 

% The GLMER model performed better than the Lasso-LR model in predicting the familiarity conditions, but not in predicting the reading goal conditions. 
% As GLMER models' addressed biases from repeated measures, both models showed better precision scores, but significantly lower recall scores, which led to also significantly lower F1 scores. 
% The differences between Lasso-LR and GLMER models can be useful for application designers to chose which measure (precision or recall) to prioritize in the final product. 

% Figure \ref{fig:ranef_slope} shows how scrolling signals can differ across participants for the same reading goal condition, depending on the usual scrolling style of a given participant. 

%%%%%%%%%%%%%%%%%%%%%%%%%%%%%%%%%%%%%%%%%%%%%%%%%%%%%%%%%%%%%%%%%%%%%%%%
\section{Discussion}
From our model analysis results, we can identify areas for future study. 
First, based on the study results, we can suggest a few personalized reading support features that may help mobile readers. 
From the ANOVA analysis (Section \ref{sec:res_anova}), we found that features based on scroll behaviors are closely related to the reading goal conditions.
Providing highlights or summaries along with an article can reduce regression scrolling and non-linear navigation behaviors in mobile reading. We hypothesize that by having an overview of an article, participants do not need to go back and forth in the text as much to integrate the content.
% making reading more linear. 
Considering touch locations and the limited screen size of a mobile phone, we noted that when touching the upper portions of the screen, a participant's hand might obscure more of the visible text. 
Reminding readers to keep a larger viewing area can help readers, especially when reading less familiar articles on mobile devices, by providing more context to readers at a time.
%We found that unfamiliar readers tended to touch the lower portion of the screen area less. Considering the limited screen size of a mobile phone, this may significantly reduces the amount of visible text during reading, as the reader's hand would block more screen area if she touch the higher locations from the screen. 

Second, more differentiated experimental settings may expose larger differences in behavioral interactions across reading conditions. 
We used different instructions to control the reading goal conditions. However, based on the post-survey results, many participants did not report reading the article differently depending on the instructions. 
One factor can be the short length of our articles (about 500 words long).
Some participants noted that they did not read the articles differently based on instructions because the articles were not too long (Section \ref{sec:post_ex_survey}). 
Similarly, the relatively low model performance for predicting familiarity conditions might be related to the nature of the articles. The articles we used in our experiments were explicitly designed to be short and to cover general topics, which means participants may not have required specific prior knowledge to understand them. 
In future studies, we can modify the experimental setting, for instance limiting the reading time or using more cognitively-demanding follow-up tasks, to see whether reading behaviors can be better differentiated by the reading condition under more realistic constraints.
% This can be caused by multiple reasons. One factor is the length of the articles. Our article stimuli were about 500 words long. 
% Similar responses for the reading goal conditions were also monitored from our preliminary analysis, even though they were given with much longer articles (2000 words). 
% As participants are more used to, or confident to general reading activities (e.g., people who are well educated, or read frequently for work), making them to employ specific reading strategy may require significant constraints in their cognitive resources. 

Third, our experimental method can be expanded to different reading scenarios. 
For this study, we used historical articles as stimuli.
In the future, we are interested in investigating reading behaviors on specialized articles, such as scientific reports or legal contracts, under more goal-driven, time-constrained, and other professional work situations. 
On the other hand, leisurely reading can evoke very different reading behaviors because the reader self-selects the material (e.g., fiction, blog articles, social network feeds) and may not have immediate reading comprehension goals in mind. 
%Articles with different purposes may also lead to different behavioral interactions of mobile readers.
Also, non-linear reading across multiple documents is a popular use case for information gathering, research, and literature search. It would be interesting to investigate how to facilitate such reading tasks in a mobile environment. 

% Based on the accumulated data from the reader, we can identify if the reader is un/familiar to the article, and identify different reading goals. 
% For example, if the reader is familiar with the article and the goal is literal reading (e.g., a top-left example of Table \ref{tab:ex_scenarios}), the system may redirect the reader into different parts of the paragraph that the reader may be more interested in than any other part of the article. If the reader is unfamiliar with the article and the goal is doing the contextual reading (e.g., a bottom-right example of Table \ref{tab:ex_scenarios}), the system may provide an interface that can promote more linear/exhaustive reading that can give the reader enough time to consume and think about newly learned ideas. 

Lastly, there are some limitations of this study. 
First, we used a 2x2 framework to model possible reading conditions, but reading behavior outside the experiment can be affected by many other factors. 
For example, engagement and motivation with the specific topic~\cite{o2016investigating} or environmental (e.g., distractions) and physiological factors (e.g., sleepiness, hunger)~\cite{schunk2012self,skinner2012developmental} can affect comprehension and learning levels. 
Identifying more comprehensive factors relating to reading conditions would be a promising area for future work. 
Second, our models for predicting the familiarity condition did not perform significantly better than the baseline. 
More exogenous predictive features might yield better prediction results. Also, we only tried a limited number of machine-learning algorithms for prediction. The results from other ML models, such as tree-based models, SVMs, or deep neural networks, might be promising candidates. 
% Although we collected data from over 300 participants, our models, especially the more complex GLMER models, experienced model convergence issues.

This paper provides an initial step to understand how different behavioral features are related to various prior conditions of mobile reading.
Future applications would gain the largest benefits from models that can make early predictions based on a small set of initial interactions, in order to efficiently customize the reading experience for the user, and help the user more quickly accomplish their reading goals. 

%, so the system can decide how to help the reader with the rest of the reading based on the early interactions. 
%Understanding the relationship between behavioral interactions across different time frames will be interesting and potentially increase the model performance by using more information. 
%Third, this study focused on predicting different reading conditions after the participant finished the reading. 
% Second, for the experiment, we used history related articles from Project Gutenberg. From the post-questionnaire, some participants expressed they enjoyed reading the articles. However, these are only particular types of the article that we thought general audiences may consume on mobile devices. 
% Other informative documents, such as news or blog articles, or longer and more expert targeted articles like research papers or xxx, can be also consumed on mobile devices. Also, non-fiction articles for leisure reading are another popular types of document that people enjoy reading on mobile devices. 
% Different types of reading may require different types of support features to improve the reader's comprehension. Future study can invest how different types of reading are related to implicit signals. 

%%%%%%%%%%%%%%%%%%%%%%%%%%%%%%%%%%%%%%%%%%%%%%%%%%%%%%%%%%%%%%%%%%%%%%%%
\section{Conclusion}
% Understanding different reading conditions would be crucial to design the personalized system for reading. 
As more people read on mobile devices, accurately predicting different reading conditions for mobile readers will be important for designing personalized support tools. 
In this paper, we presented user studies that used behavioral interaction signals to understand different reading conditions for mobile reading. We analyzed behavioral interaction data, collected from 285 MTurk participants non-invasively, and predicted different familiarity and reading goal conditions. 
Our findings suggest that features based on touch locations can be useful for distinguishing a user's familiarity with the topic. Scroll-based features and reading time are also helpful for distinguishing the reading goal (whether participants read content contextually or literally). 
We designed computational models that can predict the familiarity level conditions (56.2\%) and reading goal conditions (67.5\%) more accurately than a baseline model. 
The findings from our study can contribute to future evidence-based designs of reading support features for mobile reading applications. 
\begin{acks}
% ---masked for review---
% This research was conducted while the first author was at Adobe Research. 
\end{acks}

\bibliographystyle{ACM-Reference-Format}
\bibliography{main.bib}

\newpage
\appendix
\section{Assigning Experimental Conditions}
To keep the task of a reasonable length, each participant only read two articles in the experiment session. We could not test all four possible conditions of familiarity levels and reading goals with a single participant. Instead, we randomly chose each participant to be tested on one of the dimensions (manipulating either familiarity or reading goal) while keeping the other dimension fixed. 

\begin{table*}[h]
\caption{
Example assignments of reading conditions (Experiment 1--using articles 1 and 2).
The participant was assigned to one of the conditions in a round-robin way. 
A single fixed condition was shared during the entire experiment, while within conditions alternated between the articles in the session. 
The order of articles was counterbalanced. 
The same design was applied to Experiment 2, except using articles 3 and 4 for the experiment session.
(\texttt{Contextual} and \texttt{literal} reading goals are noted as $_C$ and $_L$. \texttt{Familiar} and \texttt{unfamiliar} conditions are noted as $^F$ and $^U$. Numbers represent the article number). 
}
\label{tab:cond_assign}
\def\arraystretch{1.3} % vertical stretch
\begin{tabular}{r l l}
\toprule
Fixed Cond.  & Within Cond. (practice) & Within Cond. (experiment) \\
\midrule
\multirow{4}{*}{$F/U$}  & $1_C, 2_L$            & $1^{F/U}_C, 2^{F/U}_L$ \\
                        & $1_L, 2_C$            & $1^{F/U}_L, 2^{F/U}_C$ \\
                        & $2_C, 1_L$            & $2^{F/U}_C, 1^{F/U}_C$ \\
                        & $2_L, 1_C$            & $2^{F/U}_L, 1^{F/U}_C$ \\ \hline
\multirow{4}{*}{$C/L$}  & $1_{C/L}, 4_{C/L}$    & $1^{F}_{C/L}, 2^{U}_{C/L}$ \\
                        & $3_{C/L}, 2_{C/L}$    & $1^{U}_{C/L}, 2^{F}_{C/L}$ \\
                        & $2_{C/L}, 3_{C/L}$    & $2^{F}_{C/L}, 1^{U}_{C/L}$ \\
                        & $4_{C/L}, 1_{C/L}$    & $2^{U}_{C/L}, 1^{F}_{C/L}$ \\
\bottomrule
\end{tabular}
\end{table*}

To reduce the number of possible combinations of articles, we preassigned the summary-article pairs for the \texttt{unfamiliar} condition. For example, for all participants who read \textit{Article 1} in the experiment session with the \texttt{unfamiliar} condition, the summary used in the practice session corresponded to \textit{Article 3}. Similarly, \textit{Article 2} was paired with \textit{Article 4}. 
When a participant began our study and submitted the initial survey questions, one of the conditions from Table \ref{tab:cond_assign} was assigned in a round-robin way. 
For example, if the participant was assigned the \texttt{familiar} condition (F) as a fixed condition, the reading goal conditions (\texttt{literal} (L) or \texttt{contextual} (C)) were selected for each article as a within condition. The order of articles was counterbalanced to reduce any ordering effects. 
\newpage
%%%%%%%%%%%%%%%%%%%%%%%%%%%%%%%%%%%%%%%%%%%%%%%%%%%%%%%%%%%%%%%%%%%%%%%%%%%%%%%%%%%%%%%%%%%%%%%%%%%%%%%%%%%%%%%%%%%%%%%%

\section{Filtering Participants}
\label{app:filtering}

We ended up using 285 participants' data (\textit{Experiment 1}: 141 participants, 77.90\% of the recruited participants; \textit{Experiment 2}: 144 participants, 75.39\%) for the analysis. Table \ref{tab:filtering} breaks down the number of participants that entered and completed the study, as well as the final number that were used for analysis. We list the number of participants filtered out due to various criteria (reading time, device name, mobile window size, and whether behavioral interactions were successfully recorded; See Sec. 5.3 in the paper). Rows of the table further break these numbers up by experimental condition to validate that the filtering did not significantly skew the number of participants left for analysis in each condition.

\begin{table*}[ht]
\caption{
The number of recruited and filtered out participants in Experiment 1 and Experiment 2. 
Four filtering criteria under \textit{No. Filtered} are not mutually exclusive with each other.
}
\label{tab:filtering}
\begin{tabular}{r p{0.55in} p{0.1in} p{0.6in} p{0.6in} p{0.6in} p{0.6in} p{0.6in} p{0.6in} p{0.6in}}
\toprule
                        & \multicolumn{2}{c}{Fixed Cond.}   & \multicolumn{3}{c}{No. Subjects}  & \multicolumn{4}{c}{No. Filtered}  \\ 
                        &&& Entered & Completed  & Analyzed & R. Time  & Dev. Name  & Wndw. Size   & Input Rec. \\
\midrule
\multirow{5}{*}{Exp 1}  &\multirow{2}{*}{Familiarity}  & F &44     &43     &40     &2  &2  &3  &3  \\
                        &                              & U &44     &42     &33     &4  &4  &5  &3  \\
                        &\multirow{2}{*}{Goal}         & C &46     &44     &36     &4  &4  &4  &2  \\
                        &                              & L &47     &41     &32     &5  &2  &4  &4  \\ 
                        &(Total)                       &   &181    &170    &141    &15 &12 &16 &12 \\\hline
\multirow{4}{*}{Exp 2}  &\multirow{2}{*}{Familiarity}  & F &47     &45     &37     &5  &2  &3  &1  \\
                        &                              & U &49     &42     &38     &3  &3  &3  &3  \\
                        &\multirow{2}{*}{Goal}         & C &48     &45     &38     &7  &5  &1  &1  \\
                        &                              & L &47     &40     &31     &2  &1  &6  &6  \\      
                        &(Total)                       &   &191    &171    &144    &17 &11 &13 &11 \\
\bottomrule
\end{tabular}
\end{table*}

\subsection{Comprehension Question Results}
\label{app:qna_response}
% The comprehension questions used for our studies were carefully normed by an English reading specialist. 
An English reading specialist carefully normed the comprehension questions used for our studies. Based on the post-experiment survey responses, participants did not find the comprehension questions too difficult. 
Despite this reporting, we noticed significant differences between the quality of responses across reading conditions. Across both familiarity conditions, participants had significantly lower response accuracy scores for the \texttt{literal} reading questions. Specifically, across the \texttt{familiar} articles, the average response accuracy for the \texttt{contextual} reading was 84.18\%, compared to 44.08\% for the \texttt{literal} reading questions (Kolmogorov-Smirnov test $p < 0.001$). Similar results were also found for the \texttt{unfamiliar} article questions (\texttt{contextual}: 83.72\% vs. \texttt{literal}: 43.29\%, Kolmogorov-Smirnov test $p < 0.001$). 
These results indicate that the difficulty of the multiple-choice questions was not equal across the familiarity conditions. Regardless, since the level of comprehension was not the scope of our study, we did not include the comprehension question results for further analysis, nor use them for data filtering (to avoid skewing the final participant numbers per condition). 
%However, it looks like the difficulty levels between different question types were not correctly controlled. 

\newpage
%%%%%%%%%%%%%%%%%%%%%%%%%%%%%%%%%%%%%%%%%%%%%%%%%%%%%%%%%%%%%%%%%%%%%%%%%%%%%%%%%%%%%%%%%%%%%%%%%%%%%%%%%%%%%%%%%%%%%%%%

\section{Initial Investigations}\label{app:pilots}
\subsection{Pilot Studies}
We ran two pilot studies to explore the possibility of using implicit interaction signals for distinguishing different types of reading. 
The goal of the first study was to investigate whether eye gaze and scroll behaviors could be used as features to differentiate types of reading. 
We recruited two participants in our lab and asked them to read articles and answer comprehension questions on a mobile phone (iPhone XR). 
Both participants were user-experience researchers. We controlled for familiarity by having them read a design article \cite{simsarian2019design} (\texttt{familiar} condition) and a machine-learning article \cite{schwartz2019green} (\texttt{unfamiliar} condition). Both articles were about 2000 words. 
During the reading, we collected participant gaze data (using a Pupil Core eye-tracker\footnote{\url{https://pupil-labs.com/products/core/}}) and behavioral interaction data (using a preliminary version of our web interface). %To collect the gaze data, we used Pupil Core eye-tracker \footnote{\url{https://pupil-labs.com/products/core/}}. 

In the second pilot study, we tested our experimental web application in a non-laboratory setting. 
We recruited 33 participants from MTurk. All participants were assigned to one of the four conditions (combinations of \texttt{familiar}/\texttt{unfamiliar} and \texttt{contextual}/\texttt{literal}) and read two articles. 
We used four shorter news articles from The Guardian (about 500 words each) \cite{vajjala2019understanding}. 
% \footnote{\url{https://github.com/nishkalavallabhi/BEA19UserstudyData}}. 
Participants spent an average of 12 minutes for the entire task and were paid \$1.50. 
For the second pilot study and the main study, we used jQuery 3.4.1, Flask 1.0.3, and SQLite to build the application for the experiment. The application was hosted on Amazon Web Service.

\subsection{Preliminary Observations}
\label{sec:preliminary_res}
From the first pilot study, we found that when reading articles on a familiar topic, participants read less linearly (or exhaustively). 
Compared to the \texttt{unfamiliar} condition, reading in the \texttt{familiar} condition showed fewer fixations and more saccades ($Chi^2$ test of independence, $p < 0.05$), and faster average saccade movement speeds (Kolmogorov-Smirnov test $p < 0.05$). 
The scrolling data from one of the participants also showed less linear reading behavior in the \texttt{familiar} condition, backed by the faster scrolling speed and larger scrolling distance (Kolmogorov-Smirnov test $p < 0.05$).
These initial results showed promise for using implicit signals to distinguish among different reading conditions. 

From the second pilot study, we found differences between touch locations across \texttt{literal} and \texttt{contextual} reading conditions, in that the former tended to initiate the scrolling from mid-horizontal and lower part of the screen (Kolmogorov-Smirnov test $p < 0.05$). This result confirmed that touch locations could be promising features for distinguishing different reading goal conditions. 
We also noticed that different participants exhibit different scrolling styles. For example, some participants made very frequent scrolls. Other participants made occasional scrolls by touching the screen less frequently and pausing more often during reading. These differences in reading behaviors motivated a within-subject design for the main experiment, in order to collect data on different reading conditions while controlling the individual participant's scrolling style.

\newpage
%%%%%%%%%%%%%%%%%%%%%%%%%%%%%%%%%%%%%%%%%%%%%%%%%%%%%%%%%%%%%%%%%%%%%%%%%%%%%%%%%%%%%%%%%%%%%%%%%%%%%%%%%%%%%%%%%%%%%%%%

\section{1-way Mixed ANOVA Results}
\label{app:anova_mixed}
We conducted a one-way mixed ANOVA analysis for each familiarity condition and reading goal condition (Table~\ref{tab:anova}). 
The ANOVA comparison was conducted with a subset of data that shares the same type of fixed condition (i.e., using familiarity level or reading goal as a fixed condition). In this way, we could compare how individual features can be distinguished by the within condition given by the fixed condition. 
As a result, we only found marginal differences between reading familiarity conditions ($p < 0.1$). 
% statistically significant differences in an article reading time feature with the reading goal conditions ($p < 0.05$). Other features for regression to instructions and vertical touch locations only found marginal differences between assigned conditions ($p < 0.1$). 

\begin{table}[h]
\caption{
One-way mixed ANOVA analysis results. We used repeated experimental conditions as mixed-effect factors in the ANOVA.
% We could find significant differences in \textit{Reading Time} feature across reading goal conditions (\texttt{contextual} vs. \texttt{literal} reading).
($^{.}$: $p < 0.1$, $^{*}$: $p < 0.05$, $^{**}$: $p < 0.01$, $^{***}$: $p < 0.001$)}
\label{tab:anova}
\begin{tabular}{r l l l}
\toprule
% Feature Type        &   Feature Name                & \multicolumn{2}{c}{2-way ANOVA}\\
Feature Type        &   Feature Name                                                       & F vs. U   & C vs. L \\
\midrule
Reading Time        &   Total article read time                                            &   &  \\
                    &   Read time per paragraph (Avg., Std.)                               & $^{.}$ (Std.)  &   \\
Scroll              &   Scroll frequency                                                   &   &   \\
                    &   Time between scroll sequences (Avg., Std.)                          &   &   \\
                    &   Travel distance per scroll sequence (Avg., Std.)                   &   &   \\
                    &   Scrolling speed per sequence (Avg., Std.)                          &   &   \\
Regression Scroll   &   Scroll frequency                                                   &   &   \\
                    &   Travel distance per scroll sequence (Avg., Std.)                   &   &   \\
                    &   Scrolling speed per sequence (Avg., Std.)                          &   &   \\
                    &   Revisiting the instruction paragraph                               & $^{.}$  &  \\
Touch Location      &   X-Axis: Ratio of touch initialization from the left or right sides &   &   \\
                    &   X-Axis: Std. of left/mid/right touch ratios                        &   &   \\
                    &   Y-Axis: Ratio of touch initialization from the low screen          & $^{.}$  &   \\
                    &   Y-Axis: Ratio of touch initialization from the mid screen          & $^{.}$  &   \\
                    &   Y-Axis: Ratio of touch initialization from the high screen         &   &   \\
                    &   Y-Axis: Std. of left/mid/high touch ratios                         &   &   \\
% TODO: add visual icons per feature type
\bottomrule
\end{tabular}
\end{table}
\newpage

%%%%%%%%%%%%%%%%%%%%%%%%%%%%%%%%%%%%%%%%%%%%%%%%%%%%%%%%%%%%%%%%%%%%%%%%%%%%%%%%%%%%%%%%%%%%%%%%%%%%%%%%%%%%%%%%%%%%%%%%
\section{Pre-screening Questionnaire Responses}
\label{app:prescreener}
We used a pre-screening questionnaire to pre-select native English speaking MTurk workers for the main study. Geographic locations of workers were limited to native English speaking countries, such as the United States, Canada, the United Kingdom, Ireland, Australia, and New Zealand. 
Overall, 1053 of 1129 respondents reported as native English speakers (93.27\%). More than 98\% of them (1110 of 1129) also reported that they are comfortable with reading English articles. The charts below summarize other question responses from MTurk workers. 
 
\newcommand\dblc{0.5}
\begin{figure}[h]
    \centering
    \begin{subfigure}{\dblc\linewidth}
        \centering\includegraphics[width=\linewidth]{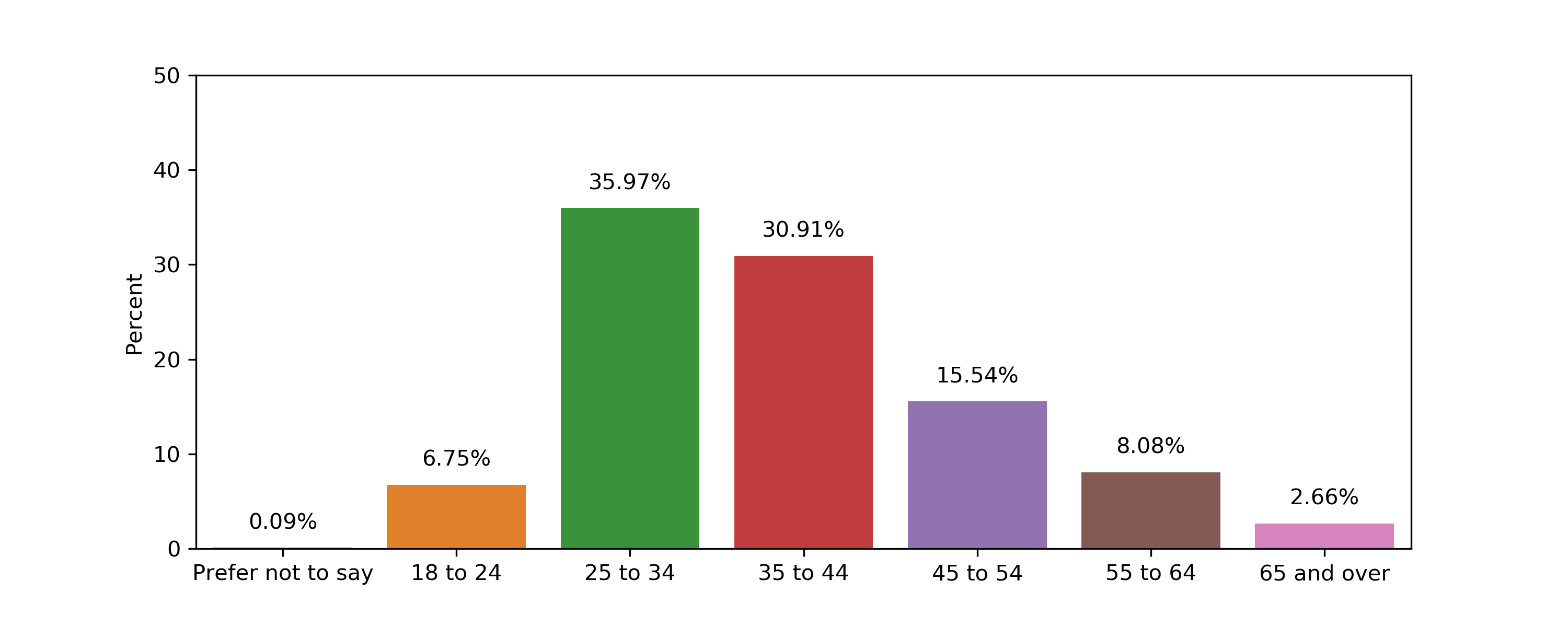}
        \caption{What is your age group?}
    \end{subfigure}~
    \begin{subfigure}{\dblc\linewidth}
        \includegraphics[width=\linewidth]{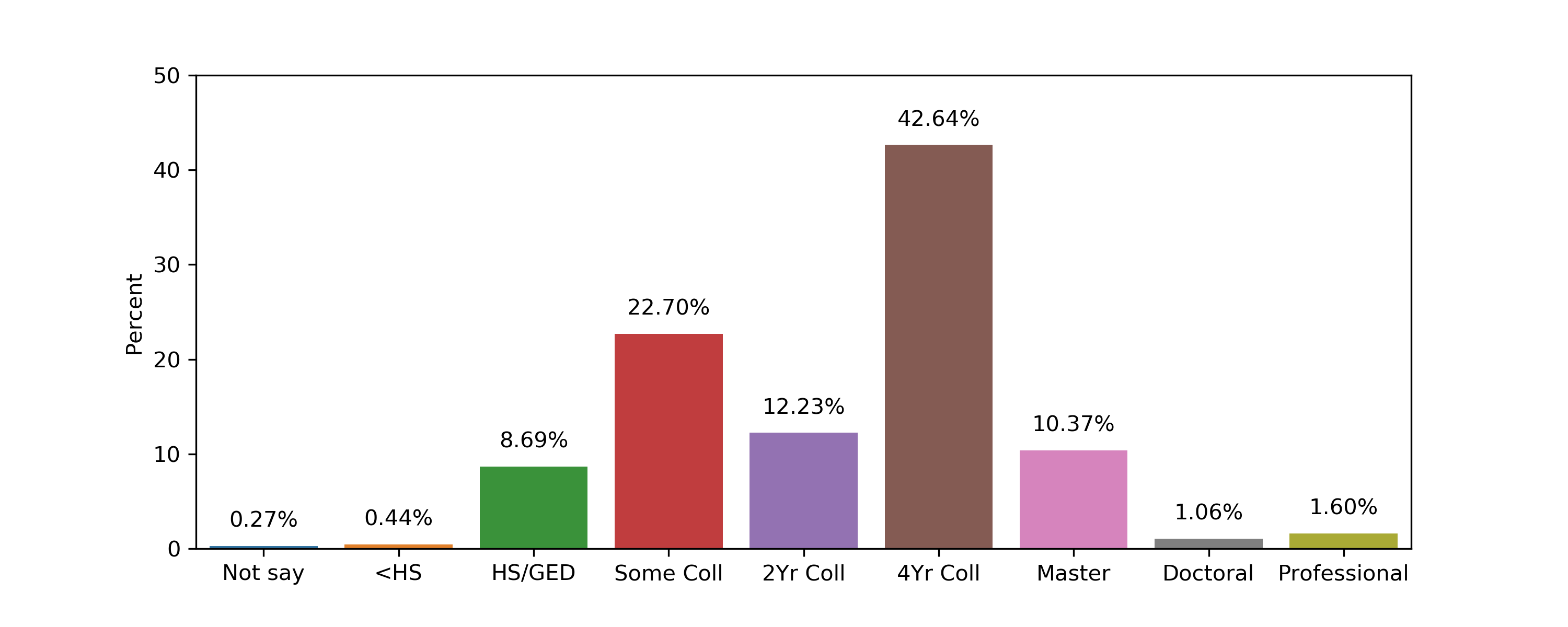}
        \caption{What is your highest education level?}
    \end{subfigure}
    
    \begin{subfigure}{\dblc\linewidth}
        \includegraphics[width=\linewidth]{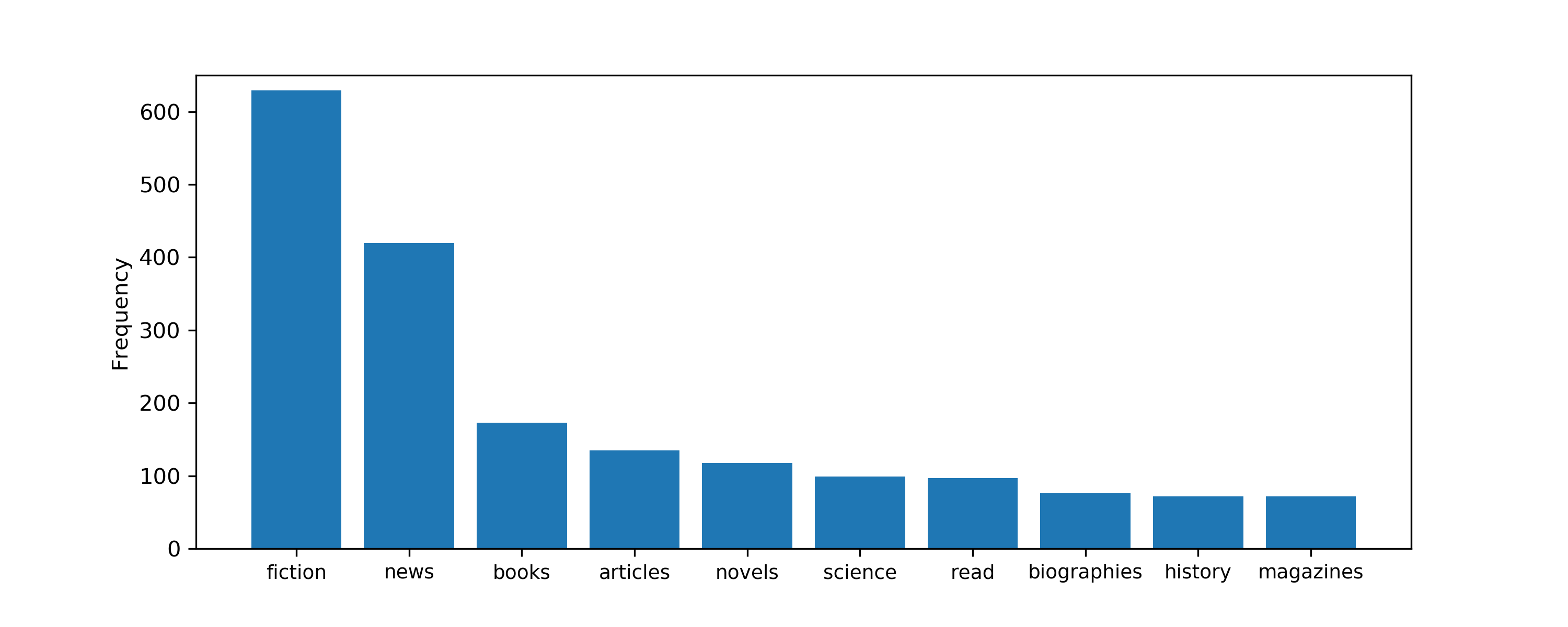}
        \caption{What do you read for leisure or personal interest? (Top 10 key words)}
    \end{subfigure}~
    \begin{subfigure}{\dblc\linewidth}
        \includegraphics[width=\linewidth]{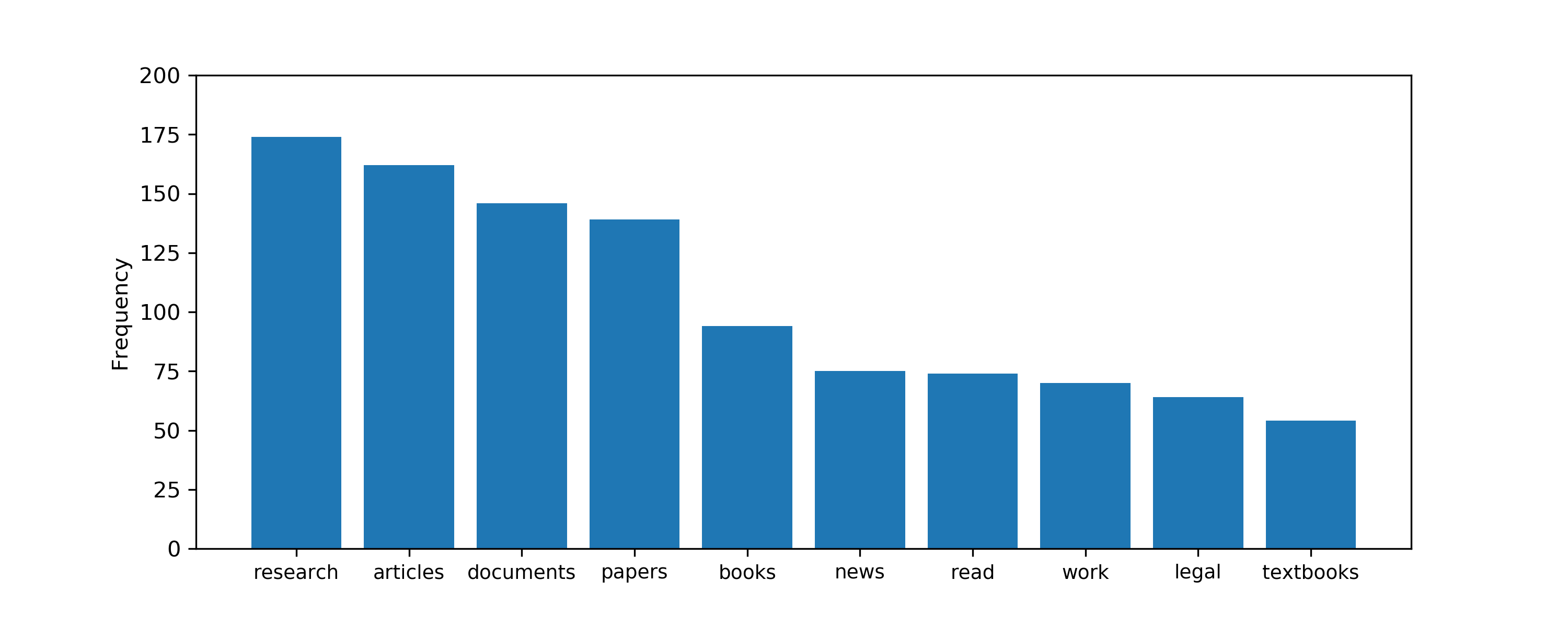}
        \caption{What do you read for work or study? (Top 10 key words)}
    \end{subfigure}    
    
    \begin{subfigure}{\dblc\linewidth}
        \includegraphics[width=\linewidth]{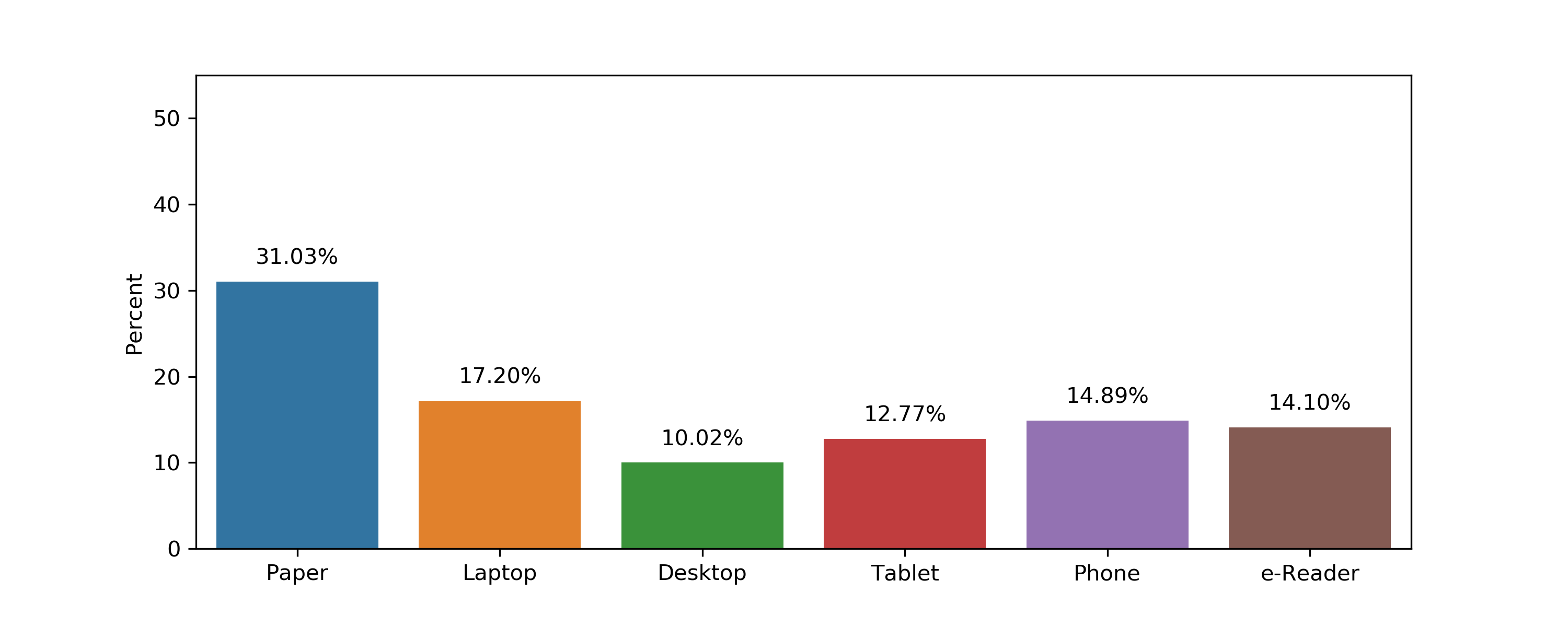}
        \caption{What device do you read on for leisure or personal interest?}
    \end{subfigure}~
    \begin{subfigure}{\dblc\linewidth}
        \includegraphics[width=\linewidth]{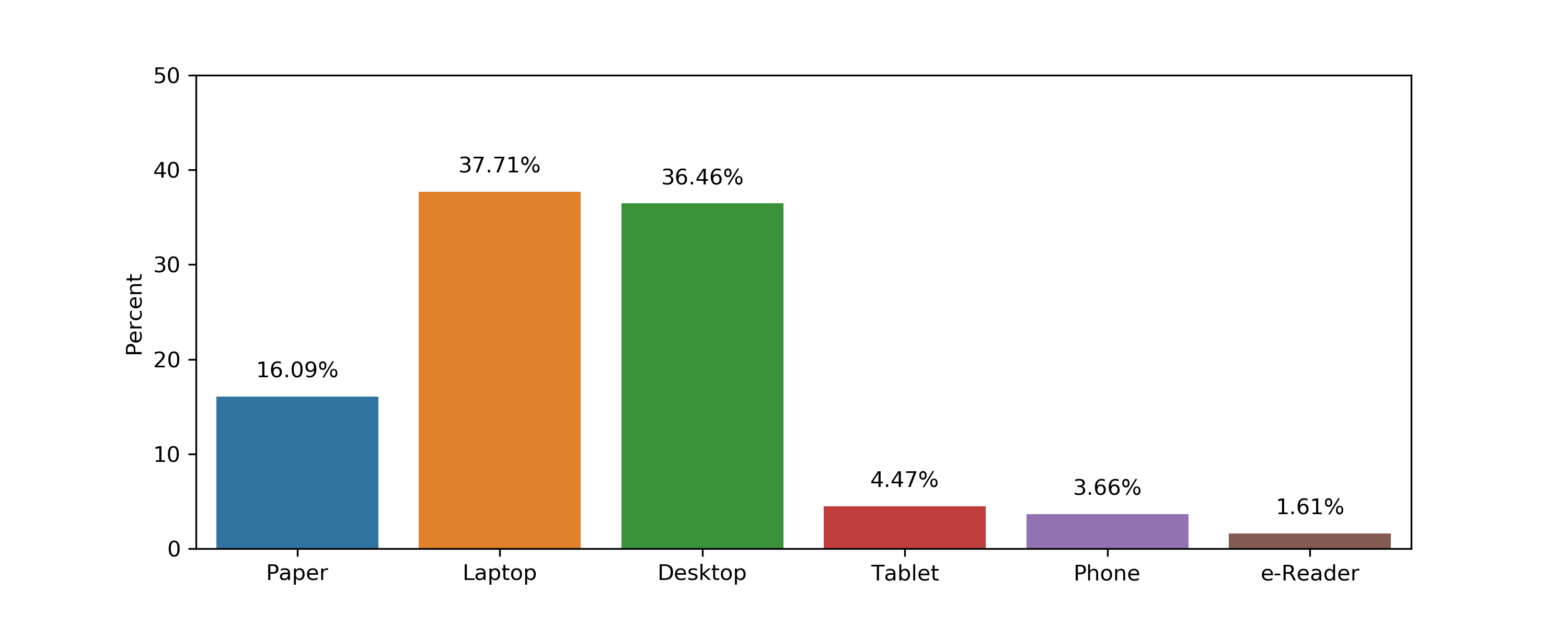}
        \caption{What device do you read on for work or study?}
    \end{subfigure}      
    
    \begin{subfigure}{\dblc\linewidth}
        \includegraphics[width=\linewidth]{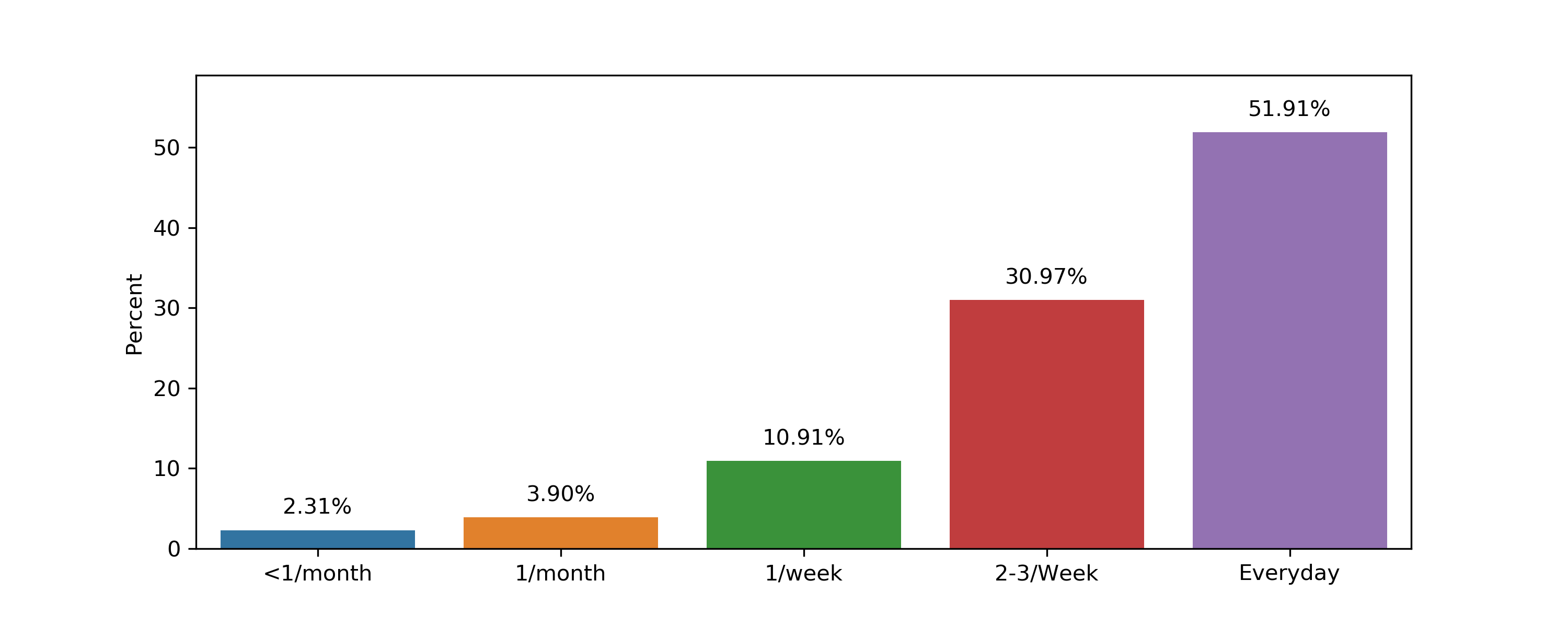}
        \caption{How often do you read English written articles for leisure or personal interest?}
    \end{subfigure}~
    \begin{subfigure}{\dblc\linewidth}
        \includegraphics[width=\linewidth]{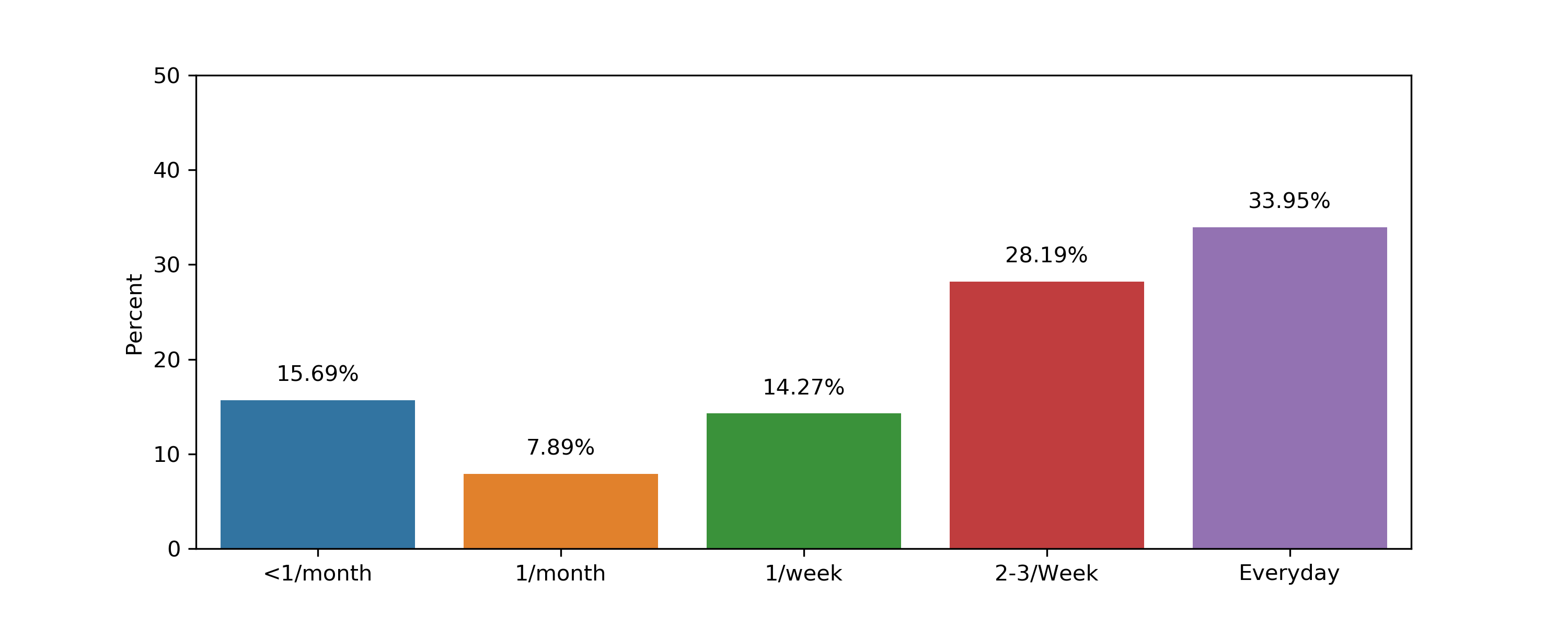}
        \caption{How often do you read English written articles for work or study?}
    \end{subfigure}        
\end{figure}
\newpage

%%%%%%%%%%%%%%%%%%%%%%%%%%%%%%%%%%%%%%%%%%%%%%%%%%%%%%%%%%%%%%%%%%%%%%%%%%%%%%%%%%%%%%%%%%%%%%%%%%%%%%%%%%%%%%%%%%%%%%%%
\section{Materials}

Here we provide the full text of all the articles, summaries, and comprehension questions that were presented to participants in our studies. As these were developed by an education expert for our studies (by modifying copyright-free articles from \url{https://www.gutenberg.org/}), we make them freely available to the research community for reuse.

\subsection{Article 1: Manhattan in the Year 1609}
\subsubsection{Practice Summary}
In 1609, Manhattan was a wild spot with beautiful trees, sand hills, and grass lands. One day, a ship from the Netherlands, called Half Moon, arrived on the shore. The ship was owned by the company of Dutch merchants. The company hired an Englishman named Henry Hudson to find a shorter course to East Indies, other than a dangerous route that go around the Cape of Good Hope. However, Hudson had not found the passage to India. Instead, he arrived at the river shore of Albany, New York.

\subsubsection{Test Article}
The long and narrow Island of Manhattan was a wild and beautiful spot in the year 1609. There were no tall houses with white walls glistening in the sunlight, no church-spires, no noisy hum of running trains, and no smoke to blot out the blue sky. Instead, there were beautiful trees with spreading branches, stretches of sand-hills, and green patches of grass. Beautiful birds and wild animals lived there in coexistence with the native people who made their houses from trees and vines. One day, the native people who lived here gathered on the shore of their island and looked on with wonder at a boat that was approaching. It was vastly different from anything they had ever seen.

The ship was called the Half Moon, and it had come all the way from Amsterdam in the Dutch Netherlands. The Netherlands was quite a small country in the northern part of Europe, not nearly as large as the State of New York. Little did they know that the Dutch people owned many lands across the globe. They had islands in the Indian Ocean, the East Indies, that were rich in spices of every kind which other European countries needed. The company of Dutch merchants who did most of the business with them was called the East India Company. They had many ships to carry out this business, and the Half Moon was one of them.

It was a long way to the East India Islands from Holland. At this time, there was no Suez Canal to separate Asia and Africa, and the ships had to go around Africa by way of the Cape of Good Hope. Besides being a long distance, it was a dangerous passage. From its name, one might think the Cape of Good Hope is a very pleasant place, but it was actually quite dangerous due to high winds and treacherous waves that were so strong they could smash ships to pieces. Therefore, the merchants of Holland, and of other countries for that matter, were constantly thinking of a shorter course to the East Indies. They knew very little about North or South America, which they believed were simply islands, and that it was possible for a passage to exist between them as a safer and more convenient shortcut to the East Indies. So, the East India Company built the ship, Half Moon, and hired an Englishman named Henry Hudson to take charge of it. Hudson was chosen because he had already made two voyages for an English company where he tried to find that same short passage between the Americas; therefore, he was supposed to know much more about it than anyone else.

When the Half Moon sailed up the river, Hudson was sure that he had found the passage to the Indies, but when the ship got as far as where Albany, NY is now, the water became shallow, and the river-banks were so close together that Hudson gave up. He declared that he had not found the passage to India, but only a river. Soon after, he returned to Holland to share his discoveries in America including the river which he called ``The River of the Mountains.''

\subsubsection{Comprehension Questions}
\paragraph{For Practice Summary}

\begin{itemize}
\item \texttt{Literal} questions
\begin{enumerate}
    \item In 1609, the island of Manhattan was covered in:
    \begin{itemize}
        \item Buildings
        \item Farms
        \item Trees
    \end{itemize}

    \item The Dutch merchants owned a: 
    \begin{itemize}
        \item Trading company
        \item Gold company
        \item Tea company
    \end{itemize}
\end{enumerate}

\item \texttt{Contextual} questions
\begin{enumerate}    
    \item The company wanted to find a shorter route to East Indies to: 
    \begin{itemize}
        \item Build a new city at Manhattan
        \item Develop a better trade route
        \item Explore the wild area
    \end{itemize}
\end{enumerate}
\end{itemize}

\paragraph{For Test Article}
\begin{itemize}
    \item \texttt{Literal} questions
\begin{enumerate}
    \item (\texttt{Literal}) The ship, Half Moon, came from:
    \begin{itemize}
        \item Germany
        \item The Netherlands
        \item Ireland
    \end{itemize}

    \item (\texttt{Literal}) The Dutch merchants came to the Americas to find:
    \begin{itemize}
        \item Spices to trade
        \item Furs to bring home
        \item A shortcut to the East Indies 
    \end{itemize}
\end{enumerate}

\item \texttt{Contextual} questions
\begin{enumerate}
    \item On their travels around the Cape of Good Hope, explorers tended to feel:
    \begin{itemize}
        \item  Excited about the journey
        \item Nervous about the rough seas
        \item Indifferent about the travel 
    \end{itemize}
    
    \item After he did not find the ocean, Hudson returned to The Netherlands feeling: 
    \begin{itemize}
        \item Disappointed
        \item Angry
        \item Ecstatic 
    \end{itemize}
\end{enumerate}
\end{itemize}

\subsection{Article 2: The Beginning of American Railroads}
\subsubsection{Practice Summary}
The United States today is largely the result of mechanical inventions, like the railroad. Before the invention, Americans crossed mountains and sailed on rivers to trade. The development of the railroad was possible from the use of mechanics for speed and the use of a smooth iron surface to eliminate friction. Its development became a solution for difficult travel and transportation in the United States. 

\subsubsection{Test Article}
The United States, as we know it today, is largely the result of mechanical inventions, and in particular of agricultural machinery and the railroad. One transformed millions of acres of uncultivated land into fertile farms, while the other allowed for transportation of crops to distant markets. Before these inventions appeared, Americans had crossed mountains, traversed through valleys, and sailed on rivers to allow for trade. 

During this time, it seemed that American tradesmen would have to rely on the internal waterways or horse and buggy to transport goods. This was an old, timely system. However, transportation changed drastically in the nineteenth century when the railroads were built. The railroads revolutionized America’s transportation system making it more efficient for the American people to travel and receive and distribute goods. It spread from coast to coast and was unlike anything that this nation had seen before. 
 
The development of the American railroad came from two fundamental ideas: the first idea was to develop speed with the use of mechanics, and the other idea was the use of a long, smooth surface to eliminate friction. These principles grew from the ancestors of the railroad track. Three hundred years before this, people used wooden rails where little “cars” were pulled to transport coal in mines. Over time, the large coal wagons drove along the public highway and made deep ruts in the road. Eventually, someone began repairing the damage by laying wooden planks in these holes. The coal wagons drove over this new roadbed so well, that certain builders started constructing special planked roadways from the mines to the river. Logs, forming what we now call ``ties,'' were placed across at intervals of three or four feet. Then, they placed thin ``rails'' of wood lengthwise. This design reduced friction and allowed cars to carry the amount of coal that two or three teams of horses had difficulty carrying. In order to preserve the road, a thin sheet of iron was laid on top of the wooden rail. 
 
Next, the overall strength and durability of the wagons needed to improve. One change was making the wagon wheels out of iron. It was not, however, until 1767, when the first rails were cast entirely of iron with a flange at one side to keep the wheel steadily in place, that the modern roadbed as it appears today came to be. 
 
The development of the railroads came as a stepping stone in technology for improving transportation. While the railroad was a familiar sight in the mining districts of England, the development of the railroad in the United States became a solution for difficult travel and transportation. It was the answer that Americans had been looking for, which stretched from sea to shining sea.

\subsubsection{Comprehension Questions}
\paragraph{For Practice Summary}

\begin{itemize}
\item \texttt{Literal} questions
\begin{enumerate}
    \item Agricultural machinery and the railroad were seen as:
    \begin{itemize}
        \item Expensive inventions
        \item Electrical inventions
        \item Mechanical inventions
    \end{itemize}

    \item The development of the railroad came from two ideas, to develop speed and to: 
    \begin{itemize}
        \item Eliminate friction on a rough surface
        \item Eliminate friction on a smooth surface
        \item Create friction for better grip
    \end{itemize}
\end{enumerate}

\item \texttt{Contextual} questions
\begin{enumerate}    
    \item Due to the development of railroads, travel most likely : 
    \begin{itemize}
        \item  Increased
        \item Stayed the same
        \item Became too demanding 
    \end{itemize}
\end{enumerate}
\end{itemize}

\paragraph{For Test Article}
\begin{itemize}
\item \texttt{Literal} questions
\begin{enumerate}
    \item Before railroads, American tradesmen had to rely on:
    \begin{itemize}
        \item  Walking on foot
        \item Internal waterways
        \item Early locomotives 
    \end{itemize}

    \item One change that was made to the wheels was constructing them out of:
    \begin{itemize}
        \item  Aluminum
        \item Metal
        \item Iron
    \end{itemize}
\end{enumerate}

\item \texttt{Contextual} questions
\begin{enumerate}
    \item Due to the development of railroads, trade most likely:
    \begin{itemize}
        \item Increased
        \item Stayed the same
        \item Became too demanding 
    \end{itemize}
    
    \item Railroad beds are mostly: 
    \begin{itemize}
        \item  The same design as its first development
        \item Somewhat different from its first creation
        \item Developing as technology advances
    \end{itemize}
\end{enumerate}
\end{itemize}

\subsection{Article 3: Water Vessels in the Pueblo Region}
\subsubsection{Practice Summary}
In the Pueblo region, water is scarce. And historically, the transportation and preservation of water was important for people in the region. Earlier people tried animal skins. Materials like gourds was used as a more effective vessel, but it was fragile. Later, people developed the water-tight basket, which was much durable. Today, these vessels considered as a tangible source of regional history.
    
\subsubsection{Test Article}
The Pueblo region of the United States is historically known as an area where pottery was practiced and worked to perfection. The architecture and arts of this region originated from the indiginous people of the desert regions of North America. Understanding the history and geography of the region gives an insight into the development of the art of ceramics and pottery.

The Pueblos’ first necessity of life is the transportation and preservation of water. Water is scarce in these regions, and can be found in small quantities, or at specific points that are extremely far away from one another. In an effort to transport water, the people once tried using skins of animals, but this was unsuccessful as the water sitting in animal skins in a hot climate would contaminate it. 
Later, a more successful water vessel was created.  
They were tubes of wood or sections of canes. 
These sections of canes were said to have been used by priests who filled them with sacred water from the ocean of the ''cave wombs'' of earth where men were born. 
Therefore, they were considered important vessels. Although these canes grew in abundance in this area, especially along the rivers, another, more effective water vessel came into play. Gourds, which also grew in abundance in these places, were better shapes and held a larger volume of water. 
The name of the gourd as a vessel is shop tom me, from shó e (canes), pó pon nai e (bladder-shaped), and tóm me (a wooden tube). 
The gourd itself is called mó thlâ â, ``hard fruit.''

While the gourd was large and convenient in its form, it was difficult to transport because of how fragile the vessel was. To help protect and transport the gourd, it was encased in a net of coarse wicker like rope which was made out of yucca leaves or of flexible splints. 

The use of this wicker with water-vessels points toward the development of the water-tight basketry of the southwest.  This explains the resemblance of many types of basketry to shapes of gourd-vessels. Eventually, these watertight baskets would inevitably replace gourd vessels. This was because, although they were difficult to manufacture, the gourds only grew in specific areas; however, the materials for basketry were everywhere. In addition, the basket vessels were much stronger and more durable. Transporting these vessels full of water at long distances were less of a danger or a hassle. Finally, because of their rough surfaces any leakages instantly stopped by a dab of mineral asphaltum, coated with sand or coarse clay externally to harden it.  Today, these vessels are not only considered antiques, but a tangible source of history showing the development of a pivotal tool over time.

\subsubsection{Comprehension Questions}
\paragraph{For Practice Summary}

\begin{itemize}
\item \texttt{Literal} questions
\begin{enumerate}
    \item Water sources tend to be described as $\rule{1cm}{0.15mm}$ in the Pueblo Region
    \begin{itemize}
        \item  Scarce
        \item Plentiful
        \item Contaminated
    \end{itemize}

    \item A more effective vessel, which held a larger volume of water, was the:
    \begin{itemize}
        \item  Gourd
        \item Animal Skin
        \item Tube
    \end{itemize}
\end{enumerate}

\item \texttt{Contextual} questions
\begin{enumerate}    
    \item Developing a good water vessel was important for: 
    \begin{itemize}
        \item  A few people
        \item Some people
        \item Many people
    \end{itemize}
\end{enumerate}
\end{itemize}

\paragraph{For Test Article}
\begin{itemize}
\item \texttt{Literal} questions
\begin{enumerate}
    \item The first vessel used to transport water was made out of:
    \begin{itemize}
        \item  Canvas
        \item Animal skin
        \item Leaves
    \end{itemize}

    \item The development of baskets resulted in the water vessel being: 
    \begin{itemize}
        \item  Unbreakable
        \item Water tight
        \item Temperature regulated 
    \end{itemize}
\end{enumerate}

\item \texttt{Contextual} questions
\begin{enumerate}
    \item The use of baskets as a water vessel were most likely:
    \begin{itemize}
        \item  Used by many people
        \item Only used by some
        \item Rarely used
    \end{itemize}
    
    \item We can infer that the development of water vessels after the creation of water tight baskets: 
    \begin{itemize}
        \item  Continued to develop
        \item Stopped developing
        \item Reverted back to use of tubes and canes
    \end{itemize}
\end{enumerate}
\end{itemize}

\subsection{Article 4: Navaho Houses}
\subsubsection{Practice Summary}
The land on the Navajo reservation is barren and void of life, and this condition affected on their houses. Building winter huts follows a set of rules and referred as ritually ``beautiful''. However, these rules do not apply for building summer huts and considered as a makeshift shelter.

\subsubsection{Test Article}
The Navajo reservation is a large area of land in the northeastern part of Arizona and the northwestern corner of New Mexico. The total area is over 11,000 square miles. About 650 square miles are in New Mexico. Unfortunately, a large part of this region consists of land that is barren and void of life. The condition of this land has had an important effect on the people, their arts, and especially their houses.

The Navaho recognize two distinct classes of hogáns—the keqaí or winter place, and the kejĭ´n, or summer place; in other words, winter huts and summer shelters. Winter huts are a staple of Navajo culture. On the outside, they resemble a mound of earth hollowed out. However, they are warm and comfortable. Their construction follows a set of rules and is considered a ritual. For example, there are ceremonies that dedicate the homes to a family before they occupy it.

Decoration on the inside or outside of the houses is uncommon, yet, the hogans are usually referred to as “beautiful.” To build this structure, strong forked timbers that are the correct length and flexibility are thrust together so that their ends properly interlock to form a cone-like frame. Stout poles lean against the apex to form the sides, and the outside is covered with bark and heaped with think dirt, forming a roomy warm interior with a level floor. To the Navajo, the house is beautiful when it is well constructed and it adheres to the ancient model.

The rules for building a regular hogán or winter house do not apply to the summer huts or shelters. The level of detail in these huts varies, but the work is done by hand and follows a specific process. This is one of the most primitive and simple shelters the Navajo builds. It starts with a center circle of greenery, generally pine or cedar. Then, it takes two men with axes a half an hour to erect one of the central circles. 

In order to start this process, a site for the hut is selected, a tree is chopped down, and the branches are trimmed from the trunk. 4 to 5 feet of branches are piled on three sides of a circle 15 or 20 feet in diameter. A fire is built in the center and blankets are thrown over outstanding branches here and there, creating a great amount of shade during the hot summer days. Although it is a makeshift shelter, it is effective to escape from the brutal heat. It is important to note that these shelters are only possible in a wooded area, and are built only to meet an emergency, like when someone is away from home and there are no hogáns in the vicinity where they can stop.
    
\subsubsection{Comprehension Questions}
\paragraph{For Practice Summary}

\begin{itemize}
\item \texttt{Literal} questions
\begin{enumerate}
    \item The land on the Navajo reservation can be characterized as:
    \begin{itemize}
        \item  Plentiful
        \item Rocky
        \item Sandy plains
    \end{itemize}

    \item Winter huts are described as: 
    \begin{itemize}
        \item  Rough
        \item Beautiful
        \item Earthy
    \end{itemize}
\end{enumerate}

\item \texttt{Contextual} questions
\begin{enumerate}    
    \item For Navajos, the type of house to build mostly relied on: 
    \begin{itemize}
        \item  Landscape
        \item Seasons
        \item Community
    \end{itemize}
\end{enumerate}
\end{itemize}

\paragraph{For Test Article}
\begin{itemize}
\item \texttt{Literal} questions
\begin{enumerate}
    \item Winter huts are built following: 
    \begin{itemize}
        \item  A set of strict rules
        \item The person who will live in the hut
        \item No rules or rituals 
    \end{itemize}

    \item Summer huts are usually:
    \begin{itemize}
        \item  Permanent homes
        \item Beautiful
        \item Makeshift
    \end{itemize}
\end{enumerate}

\item \texttt{Contextual} questions
\begin{enumerate}
    \item The rituals following the construction of winter huts are most likely: 
    \begin{itemize}
        \item  Important to the people
        \item A nuisance 
        \item Dependent on the family
    \end{itemize}
    
    \item These rituals and ways of building winter huts and summer huts are probably: 
    \begin{itemize}
        \item  Forgotten in modern times
        \item Used only when necessary
        \item A highly regarded process in the Navajo culture
    \end{itemize}
\end{enumerate}
\end{itemize}

\newpage

%%%%%%%%%%%%%%%%%%%%%%%%%%%%%%%%%%%%%%%%%%%%%%%%%%%%%%%%%%%%%%%%%%%%%%%%%%%%%%%%%%%%%%%%%%%%%%%%%%%%%%%%%%%%%%%%%%%%%%%%









\end{document}